\documentclass[prd,superscriptaddress,nofootinbib,preprint]{revtex4}
\usepackage{graphicx}
\usepackage{dcolumn}
\usepackage{bm}
\usepackage{amssymb}
\usepackage{mathrsfs}
\usepackage{amsmath}
\usepackage{epsfig}
\usepackage[dvips]{color}
\usepackage{hhline}
\usepackage[colorlinks]{hyperref}
\usepackage[normalem]{ulem}

 \usepackage{hhline} \usepackage[colorlinks]{hyperref}

\def\be{\begin{equation}}
\def\ee{\end{equation}}
\newcommand{\bea}{\begin{eqnarray}} 
\newcommand{\eea}{\end{eqnarray}}
\newcommand{\ea}{\end{array}}

\newcommand{\dms}{\mbox{$\Delta m^2_{sol}$}}
\newcommand{\dma}{\mbox{$\Delta m^2_{atm}$}}

\newcommand{\nonubb}{\mbox{$ 0\nu\beta\beta $ }}

\begin{document}
\title{Neutrinoless 
double-$\beta$ decay in TeV scale Left-Right symmetric models}
 \author{Joydeep Chakrabortty} 
 \email{joydeep@prl.res.in} 
\affiliation{Physical Research Laboratory, Ahmedabad-380009, India}
 \author{H. Zeen Devi}
 \email{zeen@prl.res.in}
 \affiliation{Physical Research Laboratory, Ahmedabad-380009, India}
\author{Srubabati Goswami}
 \email{sruba@prl.res.in}
 \affiliation{Physical Research Laboratory, Ahmedabad-380009, India}
 \author{Sudhanwa Patra}
\email{sudha.astro@gmail.com}
 \affiliation{Institute of Physics, Bhubaneswar-, India}

\begin{abstract}
In this paper we study in detail the 
neutrinoless double beta decay in left-right symmetric models with 
right-handed gauge bosons at TeV scale which is within the presently 
accessible reach of colliders.  We  discuss the different diagrams that 
can contribute to this process and identify the dominant ones for the 
case where the right-handed neutrino is also at the TeV scale. 
We calculate the contribution to the effective mass governing neutrinoless
double beta decay assuming type-I, and type-II dominance and discuss what 
are the changes in the effective mass due to the additional contributions. 
We also discuss the effect of the recent Daya-Bay and RENO measurements 
on $\sin^2\theta_{13}$ on the effective mass in different scenarios.  
\end{abstract}
\maketitle
\section{Introduction }

Neutrino oscillation experiments have steadfastly established
that neutrinos are massive and mix between different flavours. 
The Standard Model (SM) does not accommodate any right-handed (RH) partner for the
neutrino and hence they remain massless. Thus, existence of neutrino mass 
requires one to tread beyond the standard paradigm.
A prevalent way to generate small neutrino mass is through
the seesaw mechanism.
This is due the Weinberg operator
$\kappa LLHH$ \cite{Weinberg:1979sa} where L and H are respectively lepton and Higgs fields
transforming as $SU(2)$ doublets;
$\kappa$ is the effective coupling which
has inverse mass dimension.
Such terms violate lepton number by 2 units
and hence essentially predict neutrinos to be Majorana particles.
Violation of lepton number  can be manifested in 
neutrinoless double beta decay ($\nonubb$) process : 
\begin{center} 
$(A,Z) \rightarrow (A, Z+2) + 2 e^{-}$
\end{center}  
which is therefore of prime importance as it can ascertain the 
nature of the neutrinos and hence the mechanism of neutrino mass generation. 
The best limit on the half life of this process at present is  
$T_{1/2} < 3 \times 10^{25}$ years coming from the 
Heidelberg-Moscow experiment using $^{76}{Ge}$ \cite{KlapdorKleingrothaus:2003md}.
This translates to a bound on the effective neutrino mass 
\cite{Rodejohann:2011mu} 
\be
m_{\mathrm{eff}} \leq 0.21 - 0.53~ {\rm eV} 
\ee 
where the range is due to the uncertainty in the nuclear matrix elements. 
There are many 
other upcoming experiments to improve this bound \cite{0nubbexpt} 
and also to 
test the claim of the positive evidence for  $\nonubb$ 
by a part of the Heidelberg-Moscow Collaboration \cite{KlapdorKleingrothaus:2006ff,KlapdorKleingrothaus:2004wj}. 

If the only contribution to $\nonubb$ is through exchange of
light neutrinos 
then the effective neutrino mass is just 
the absolute value of the 
($ee$) element of the low energy neutrino mass matrix in the flavour 
basis.  
However there can be  
other scenarios 
like models with heavy right-handed neutrinos
\cite{Mitra:2011qr,Ibarra:2011xn,Pascoli:2007qh} 
, R-parity violating Supersymmetry 
\cite{Hirsch:1995ek, Hirsch:1995zi, 
Faessler:1997db, Pas:1998nn}, extra 
dimensional \cite{ED-0nnb} scenarios etc. which can give rise
to  additional 
diagrams contributing to the neutrinoless double beta decay process.  
Of special importance are the scenarios 
where the scale of new physics is at $\sim$ TeV since then 
the additional contributions to $\nonubb$ can be significant 
\cite{Allanach:2009iv,Tello:2010am,Nemevsek:2011aa}.
Such models have recently gained prominence since it 
these can be probed at the LHC and can also 
give rise to Lepton Flavour Violating processes\cite{LFV-0nnb}. 
It is noteworthy to mention at this point that lepton number violation
can also be probed at LHC through the so called golden process \cite{Keung:1983uu} producing
same sign di-leptons in the final state 
and complimentary 
information from LHC and $\nonubb$ may help in consolidating
the nature of new physics and origin of neutrino mass. 

In this paper we focus on the  TeV scale left-right symmetric seesaw models 
for generation of neutrino mass and the implications for $\nonubb$.
Left-right (LR) symmetric models  were motivated from the perspective of 
restoring  parity symmetry at a high scale. 
This required placing
the left- and right- handed fermions on the same footing
by putting them as part of $SU(2)_L$ and $SU(2)_R$ doublets
respectively \cite{LR}. The observation of parity violation in weak 
interaction indicates that the LR symmetry is broken at a low scale and 
in the Minimal LR symmetric model this is achieved by triplet scalar 
fields.   Such a framework naturally generates neutrino masses 
through type-I seesaw due to the right-handed neutrinos \cite{seesaw1}  
and type-II seesaw via the triplet scalars \cite{seesaw2}.     
 
Because $SU(2)_R$ is a gauge symmetry  
the LR symmetric models contain right-handed charged currents mediated
by $W_R$ boson.
If the right-handed $W_R$ masses are around TeV then it is
possible to access them at LHC.  
This has motivated several studies
recently with the right-handed $W_R$ at the TeV scale 
\cite{lrtev,bhupal1,jd-gluza}.
In such a scenario, 
one can have  contributions 
to $\nonubb$ 
from both left-handed (LH) and right-handed currents via exchange of 
light and heavy neutrinos respectively. 
Additionally there can be diagrams mediated by the charged Higgs fields. 
Our work scrutinizes in detail the relative contributions of various 
diagrams for TeV scale LR symmetric models. 
We consider seesaw scenarios with type-I, and type-II dominance 
and discuss the behaviour of the effective mass in different 
limits.  The implications for type-II seesaw dominance in the context 
of LR symmetric models have been considered in 
\cite{Tello:2010am}. 
In our work we present the analytic expressions of the effective mass 
in different limits and discuss their dependence on the 
neutrino oscillation parameters. In particular we incorporate the 
recent results on $\theta_{13}$ measurement from Daya-Bay 
\cite{An:2012eh} and RENO \cite{RENO} experiment and discuss the consequences. 
We also study the contribution of the triplet Higgs mediated diagrams and 
discuss for which scenarios it can contribute.  
     
The plan of the paper is as follows. In the next section we discuss 
type-I, and type-II seesaw in LR symmetric model. 
In section 3  we consider neutrinoless double beta decay due to light 
neutrinos. 
In section 4 we examine the other 
additional contributions to $\nonubb$ within the LR model in both type I, and type II dominance. 
We also summarize briefly the possible impact of the Higgs triplet contribution at the end of this section. 
In the Appendix we present the Feynman diagrams and estimated different contributions. 
We conclude with an overview of our study. 


\section{Neutrino mass in LR symmetric model}

In  left-right symmetric models, the standard model gauge group is  
augmented to include a right-handed $SU(2)$ counterpart enlarging the group to  
$SU(3)_{c}\otimes SU(2)_{L}\otimes SU(2)_{R}\otimes U(1)_{B-L}$.
Thus along with the left-handed doublet fermions we also have 
replicas that transform as doublets under $SU(2)_R$.
The $SU(2)_R$ is related to the  
$SU(2)_L$, by a discrete symmetry 
which is the anthem of the left-right symmetric model.
We restrict ourselves within the so-called minimal form of the model
that includes 
a bidoublet($\Phi$), triplet($\Delta_{L/R}$)-scalars, 
right-handed neutrino($N_R$),
extra gauge bosons($W_R^{\pm},Z_2$) along with the SM particles.
The assigned quantum numbers of these extra scalars under $SU(2)_{L}\otimes SU(2)_{R}\otimes U(1)_{B-L}$ read as:\\
\begin{equation}
\Phi \equiv (2,2,0), \Delta_L \equiv (3,1,1), \Delta_R \equiv (1,3,1).
\end{equation}
  
The Lagrangian giving the neutral fermion mass terms 
is 
\begin{equation} 
{\cal L}_{Y} =   f_L l_L^T C i\sigma_2 \Delta_L  l_L + f_R l_R^T C i\sigma_2 \Delta_R l_R 
                  +\bar{l}_R (y_D \phi + y_L \tilde{\Phi}) l_L + h.c.,
\label{LR-triplet+bi}
\end{equation}  
where $l_L (l_R)$ denotes the left(right)- handed fermion doublets 
and $\tilde{\Phi}=  \sigma_2 \Phi^* \sigma_2$. 
The bi-doublet Higgs acquires vacuum expectation values (vev) as:
\be
<\Phi>=\left(\begin{array}{cc} 
v & 0 \\ 
0 & v^{'} 
\end{array} \right) .
\label{vevPhi}
\ee
The charge lepton masses are given as: 
\begin{eqnarray}
 m_l &=& y_D v^{'} + y_L v\\
m_D &=& y_L v^{'} + y_D v.
\end{eqnarray}
Here, we consider $y_D v >> y_L v^{'}$, and $y_L v >> y_D v^{'}$ so that 
the Yukawa coupling matrix responsible for charged fermion masses is 
$y_L$ while for the neutrino masses 
it is $y_D$.  

Once the Higgs fields develop vev
the neutral fermion mass matrix becomes 
\be
M_\nu \equiv \left(\begin{array}{cc} 
f_Lv_L & y_D v \\ 
y_D^T v & f_R v_R 
\end{array} \right) ,
\label{eq:LR-seesawfull}
\ee
where $<\Delta_L>=v_L$,  $<\Delta_R>=v_R$.   
After block diagonalizing $M_\nu$ in the seesaw approximation 
($f_R v_R >> y_D v$) 
we get 
\begin{equation}
 (m_{\nu}^{light})_{3\times3} = f_Lv_L + \frac{v^2}{v_R}y_D^T f_R^{-1} y_D, 
\label{mnulight}
\end{equation}
\begin{equation} 
(m_{R}^{heavy})_{3\times3}  =  f_R v_R,
\label{mnuheavy} 
\end{equation}
where, the first term in Eq.(\ref{mnulight}) is due to type-II seesaw  
whereas the second term corresponds to type-I seesaw mediated by 
right-handed neutrinos.  

In this model the Yukawa couplings $f_L$ and $f_R$ are not independent but related to each other.
We can have either $f_L = f_R$, and $U_L = U_R$ or $f_L = f_R^{*}$, and $U_L = U_R^{*}$
as an artifact of the LR-symmetry. 
This constrains the extra freedom in the Yukawa sector and reduces the 
number of free parameters.

\section{Neutrinoless double-$\beta$ decay in three generation picture }

In the standard three generation picture the time period for 
neutrinoless double beta decay is given as, 
\begin{equation}
\frac{\Gamma_{  \nonubb }}{\text{ln\,2}} = G 
\left|\frac{\mathcal{M}_{\nu}}{m_e}\right|^2      
|m_\nu^{ee}|^2  ,
\end{equation}
where $G$ contains the phase space factors, $m_e$ is the electron mass,  
${\mathcal{M}_{\nu}}$ is the nuclear matrix element. 
\begin{equation} 
|m_\nu^{ee}| = |U^2_{ei} \, m_i|, 
\label{effectiveml}
\end{equation} 
is the effective neutrino mass that appear in the 
expression for time period for  neutrinoless double beta 
decay.   

The unitary matrix $U$ is the so called PMNS mixing 
matrix which coincides with the neutrino mixing matrix in the 
basis where charged lepton mass matrix is diagonal. 
The standard parametrization for this is 
\begin{equation}
U   =  \left(
 \begin{array}{ccc}
 c_{12} \, c_{13} & s_{12}\, c_{13} & s_{13}\, e^{-i \delta}\\
 -c_{23}\, s_{12}-s_{23}\, s_{13}\, c_{12}\, e^{i \delta} &
 c_{23}\, c_{12}-s_{23}\, s_{13}\, s_{12}\,
e^{i \delta} & s_{23}\, c_{13}\\
 s_{23}\, s_{12}-\, c_{23}\, s_{13}\, c_{12}\, e^{i \delta} &
 -s_{23}\, c_{12}-c_{23}\, s_{13}\, s_{12}\,
e^{i \delta} & c_{23}\, c_{13}
 \end{array}
 \right) P \, .
\label{upmns}
\end{equation}
The abbreviations used above are
$c_{ij} = \cos \theta_{ij}$, $s_{ij} = \sin \theta_{ij}$,
$\delta$ is the Dirac CP phase while the phase matrix 
$P = {\rm diag}(1, e^{i \alpha_2}, e^{i (\alpha_3 + \delta)})$
contains the Majorana phases $\alpha_2$ and $\alpha_3$. 
The 3 $\sigma$ ranges of the mass squared differences and mixing angles 
from  global analysis of oscillation data are depicted in Table \ref{table-osc}. 
\begin{table}[htb!]
\begin{tabular}{ccc}\\
        \hline
        parameter & best-fit & 3$\sigma$\\
        \hline
        $\Delta m^2_{\rm {sol}} [10^{-5} \mbox{eV}^2]$ & 7.58 & 6.99-8.18\\
        $|\Delta m^2_{\rm {atm}}| [10^{-3} \mbox{eV}^2]$ & 2.35 &2.06-2.67\\
        $\sin^2\theta_{12}$ & 0.306 & 0.259-0.359 \\
        $\sin^2\theta_{23}$ & 0.42 & 0.34-0.64\\
        $\sin^2\theta_{13}$ & 0.021 & 0.001-0.044\\
        \hline
\end{tabular}
\caption{The best-fit and 3$\sigma$ values for the mass squared differences and mixing angles from \cite{Fogli:2011qn}.}
\label{table-osc}
\end{table}
Recently the evidence 
of non-zero $\theta_{13}$ at  more than 5$\sigma$  were 
reported by the Daya-Bay \cite{An:2012eh} and RENO \cite{RENO} experiments
with best-fit values and 3$\sigma$ ranges of $\sin^2\theta_{13}$ as, 
$$ {\mathrm {Daya-Bay}}:~~~ \sin^2\theta_{13} = 0.023~ (\mathrm{best-fit}); 
~~0.009 - 0.037~ (3\sigma~~ \mathrm{range}) 
$$ 
$$
{\mathrm {RENO}}:~~~ \sin^2\theta_{13} = 0.026~ (\mathrm{best-fit}); 
~~0.015- 0.041~ (3\sigma~~ \mathrm{range}).$$ 

Thus there is an increase in the lower limit and a decrease in the upper limit of $\theta_{13}$ as compared to
the values given in Table \ref{table-osc}. 

With the  parametrization of the mixing matrix in Eq.(\ref{upmns}) 
the effective mass  
$|m_\nu^{ee}|$ becomes, 
\begin{equation} 
|m_\nu^{ee}| = \big |m_1 c_{12}^2 c_{13}^2 + m_2 s_{12}^2 c_{13}^2 e^{ 2 i \alpha_2} 
+ m_3 s_{13}^2 e^{2 i \alpha_3} \big|. 
\label{mnuee-effective}
\end{equation}
The effective mass  
assumes different values depending on whether 
the neutrino mass states follow normal or inverted hierarchy 
or they are quasi-degenerate. 
Where, 
\begin{itemize} 
\item
Normal hierarchy (NH)  refers to the arrangement which corresponds to
$ m_1 < m_2 << m_3$ with
\begin{equation}
 m_2 = \sqrt{m_1^2 +\Delta m_{\rm sol}^2}\;,\qquad 
 m_3 = \sqrt{m_1^2 +\Delta m_{\rm atm}^2 + \Delta m_{\rm sol}^2}\;. 
\label{NH}
\end{equation}

\item
Inverted hierarchy (IH) implies $m_3 << m_1 \sim m_2$ with
\begin{equation}
 m_1 = \sqrt{m_3^2 +\Delta m_{\rm atm}^2}\;,\qquad 
 m_2 = \sqrt{m_3^2 +\Delta m_{\rm sol}^2 +\Delta m_{\rm atm}^2 }\; \label{eq1:IH_m1_m2}.
\end{equation}

\item 
Quasi degenerate neutrinos correspond to
$m_1 \approx m_2 \approx m_3 >> \sqrt{\Delta m^2_{atm}}$.

\end{itemize} 

Fig.(\ref{lightneutrino_general}) displays the effective mass 
governing $\nonubb$  as a function of the lowest mass scale 
in the standard three generation picture for various mass schemes. 
The gray (lighter)
band  for NH corresponds to varying the parameters in their 
3$\sigma$ range as given in Table \ref{table-osc} 
whereas the black band corresponds to the best-fit parameters.
In both figures the Majorana phases are varied between 0 to 2$\pi$.
The figure in the left panel is for 3$\sigma$ ranges of $\sin^2\theta_{13}$ 
from Table \ref{table-osc} whereas the plots in the right panel use 3$\sigma$ range 
of $\sin^2 \theta_{13}$ as
measured by the Daya-Bay experiment. 
Below we discuss briefly the  limiting values of 
effective mass for different mass schemes. 
We also comment on the  impact of the Daya-Bay  and RENO measurement 
of $\theta_{13}$ on the effective mass. 
The  different limits of Eq.(\ref{mnuee-effective}) depend on the relative 
magnitudes of
$m_1$, $\sqrt{\Delta m^2_{\mathrm {sol}}} \sim 9 \times 10^{-3}$ eV,
$\sqrt{\Delta m^2_{atm}} \sim 0.05 $ eV.
Of special importance is the mass ratio 
$r= \big | \frac{\Delta m^2_{sol}}{\Delta m^2_{atm}} \big |$.
In Table \ref{table-cancellation} we put the 3$\sigma$ ranges of  different 
parameters  and their combinations that will be relevant 
for our discussion. 

\begin{table}[htb!]
\begin{tabular}{|c|c|c|c|c|c|}
\hline \hline
   &  $\sqrt{r}$ & $\sqrt{r} \sin^2 \theta_{12}$  & $\sqrt{r} \cos 2\theta_{12}$  & $\tan^2 \theta_{13}$ & $\sqrt{r} \tan^2 \theta_{13}$  \\
\hline
Maximum  & 0.2 & .072 &  .096 & .046 (.037) & 0.009 (0.007)\\
\hline
Minimum & 0.16 & .042 & .046 & .001 (.009) & 0.0001 (0.002)\\
\hline
\end{tabular}
\caption{The 3$\sigma$ ranges of different combinations of oscillation
parameters relevant for understanding the behaviour of the effective mass 
in different limits. The values in the parentheses in the last two columns are
from Daya-Bay results.} 
\label{table-cancellation}
\end{table}
 
\subsection{\textcolor{blue}{Normal Hierarchy}}
\begin{figure}[htb]
\begin{minipage}[t]{0.48\textwidth}
\hspace{-0.4cm}
\begin{center}
\includegraphics[width=5cm,angle=-90]{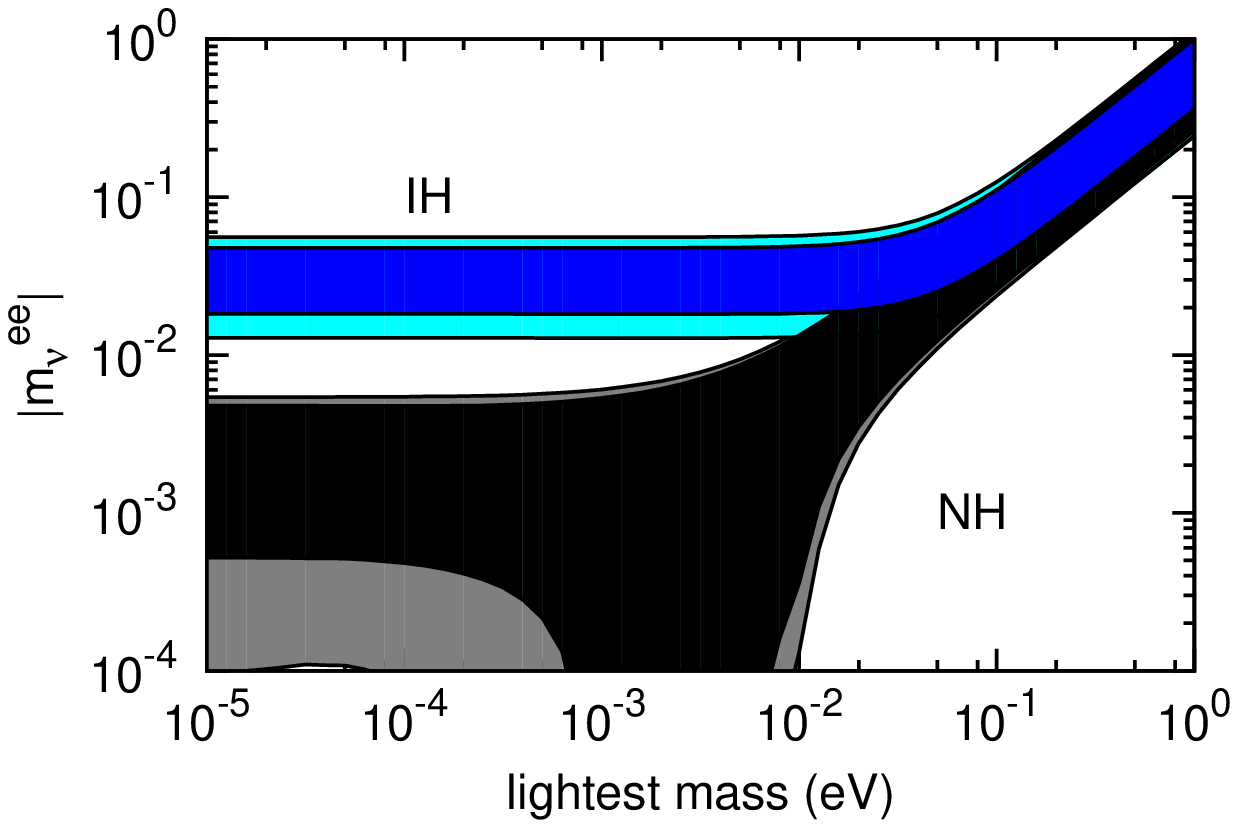}
\end{center}
 \end{minipage}
 \hfill
 \begin{minipage}[t]{0.48\textwidth}
 \hspace{-0.4cm}
 \begin{center}
 \includegraphics[width=5cm,angle=-90]{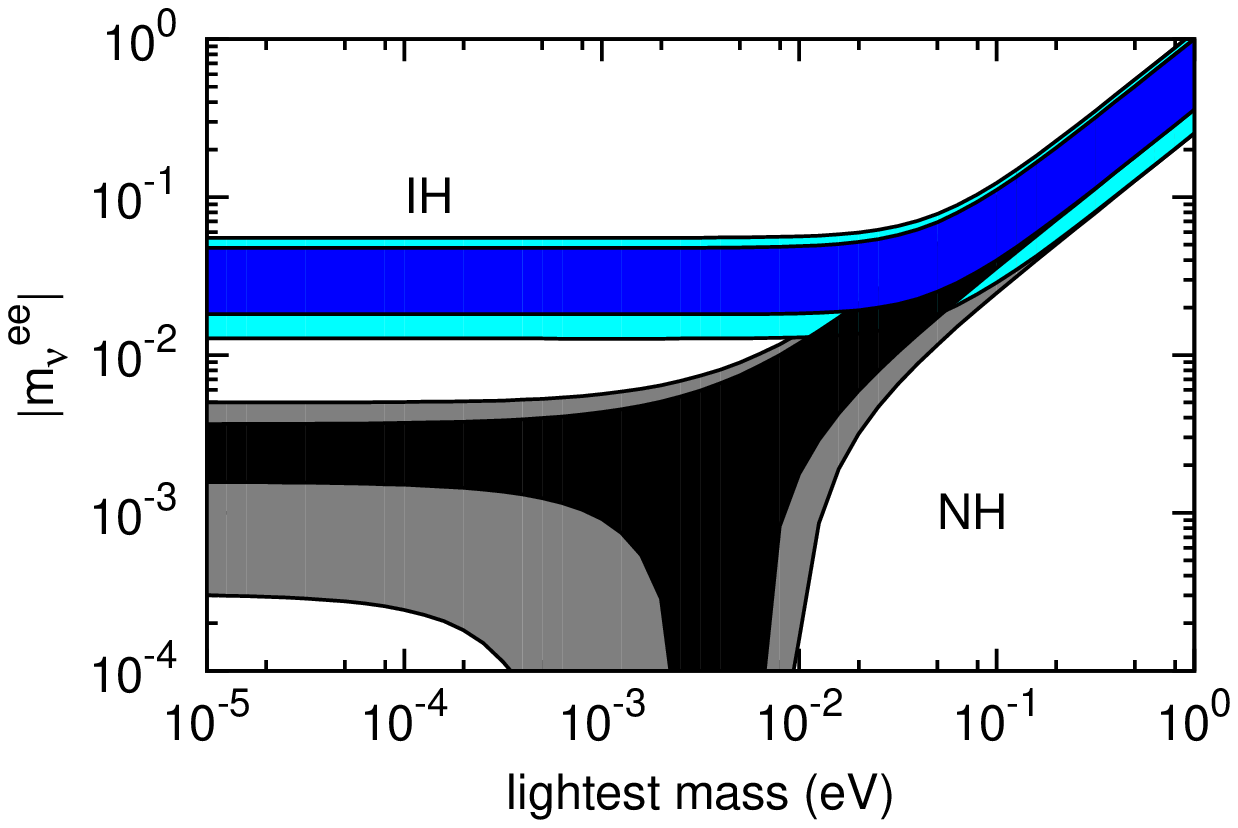}
 \end{center}
 \end{minipage}
 \caption{The canonical contribution 
from light neutrino mass with $\theta_{13}$ from \cite{An:2012eh}(left) and from \cite{Fogli:2011qn}(right)
to the neutrinoless double beta decay.}
\label{lightneutrino_general}
 \end{figure}

In the strictly hierarchical regime
($ m_1 << m_2 \approx \sqrt{\Delta m^2_{\mathrm sol}} << m_3 \approx \sqrt{\Delta m^2_{atm}}$) 
the effective mass  due to the light neutrinos, can be
approximated by
\begin{equation} 
|m_\nu^{ee}|_{NH} = \sqrt{\Delta m^2_{atm}}~  
\big |\sqrt{r} s_{12}^2 c_{13}^2 e^{ 2 i \alpha_2} + s_{13}^2 e^{2 i \alpha_3} \big |. 
\label{mnuee_xnh} 
\end{equation}
The maximum value of this  
corresponds to the phase-choice $\alpha_2 = \alpha_3 = 0$ while the minimum 
occurs for   $\alpha_2 =0, \alpha_3 = \pi$.
Since the upper limit of $\sin^2\theta_{13}$ 
from Daya-Bay measurement is less than the upper limit of $\sin^2\theta_{13}$
given in Table  \ref{table-osc} the maximum value  of $|m_\nu^{ee}|_{NH}$ 
becomes lower and minimum value becomes higher
in the second panel of Fig.(\ref{lightneutrino_general}). 
In this limit both the terms can be comparable resulting in
complete cancellation if the following condition is fulfilled: 
\be
\sqrt{r}~sin^ 2\theta_{12} = \tan^2\theta_{13}.
\ee 
Comparing the columns 3 and 5 of Table \ref{table-cancellation} we see that  
using the range of $\theta_{13}$  measured by Daya-Bay experiment, complete cancellation 
is not satisfied exactly. 
This condition changes as we increase $m_1$ and approach
the limit of partial hierarchy:  
$m_1$ 
$\approx m_2 \approx \sqrt{\Delta m^2_{sol}} << m_3 \approx \sqrt{\Delta m^2_{\rm atm}}$
the minimum value for the above expression correspond to the phase choices 
$\alpha_2 = \alpha_3 = \pi/2$,
The condition for complete cancellation now alters to,   
\be
\sqrt{r}~cos 2\theta_{12} = \tan^2\theta_{13}.
\ee
From the 3$\sigma$ ranges of parameters in Table \ref{table-cancellation} 
we see that in this region complete cancellation takes place in both panels in 
Fig.(\ref{lightneutrino_general}) resulting in very low values of 
the effective mass. 
Thus, maximum and  minimum values of the effective mass   
is sensitive to the value of $\theta_{13}$  in this region. 

\subsection{\textcolor{blue}{Inverted Hierarchy}}

In the inverted hierarchical case, $m_1 \approx m_2 >> m_3$ and 
for the smaller values of $m_3$ such that $m_3  
<< \sqrt{\Delta m^2_{\rm atm}}$,
$m_2 \approx m_1 \approx \sqrt{\Delta m^2_{\rm atm}}$ 
the effective mass is given as
\begin{equation}
\big |m^{ee}_{\nu}\big|_{IH} =\sqrt{\Delta m^2_{atm}}(c_{12}^2 c_{13}^2 + s_{12}^2 c_{13}^2 e^{2i\alpha_2}). 
\end{equation}

$\big |m^{ee}_{\nu}\big|$ in this region is independent of
$m_3$ and is bounded from above and below by a maximum and minimum value 
given by 
\be
\big |m^{ee}_{\nu}\big|_{max}=c_{13}^2\sqrt{\Delta m^2_{atm}} 
~~(\mathrm{\alpha_2=0, \pi}) 
\label{c-9}
\ee
\be
\big |m^{ee}_{\nu}\big|_{min}= c_{13}^2 cos\,2\theta_{12}\sqrt{\Delta m^2_{atm}}\\
~~(\mathrm{\alpha_2=\pi/2,~ 3\pi/2}).
\label{c-10}
\ee
As $m_3$ approaches $\sqrt{\Delta m^2_{\mathrm atm}}$, the other two masses 
can be approximated as, 
$m_2 \approx m_1 \approx \sqrt{2 \Delta m^2_{\rm atm}}$  
and the effective mass becomes, 

\begin{equation}
\big |m^{ee}_{\nu}\big|_{IH} =\sqrt{\Delta m^2_{atm}}\Big(\sqrt{2} c_{13}^2 (c_{12}^2 + s_{12}^2 e^{2i\alpha_2}) + s_{13}^2 e^{2i\alpha_3}\Big)
\end{equation}
The maximum value of this corresponds to $\alpha_2 = \alpha_3 =0$ and complete
cancellation cannot take place. 
For still higher values 
of $m_3$ one transcends into the quasi degenerate limit. 

\subsection{\textcolor{blue}{Quasi Degenerate} } 

In this regime the effective mass (for  both normal and inverted ordering)
is given as
\begin{equation} 
|m_\nu^{ee}|_{QD}
=  m_0 \big|c_{12}^2 c_{13}^2 + s_{12}^2 c_{13}^2 e^{2 i \alpha_2} 
+ s_{13}^2 e^{2 i \alpha_3} \big|, 
\end{equation}
where $m_0\approx m_1 \approx m_2 \approx m_3$.
Thus the effective mass  due to the light neutrinos
increases linearly as the common mass scale.

These well known features   are reflected in 
Fig.(\ref{lightneutrino_general}).
In particular the Fig.(\ref{lightneutrino_general})(right) shows the impact of 
Daya-Bay observations.  This impact seems to be nominal for IH. 
For NH it affects the low $m_1$ region.

\section{Neutrinoless double-$\beta$ decay in minimal LR model }

If we consider the left-right symmetric model then there can be several
additional diagrams contributing to this process 
\cite{Hirsch:1996qw}:\\
(a) The  right-handed current mediated by $W_R$ can contribute to the 
process 
through the  exchange  of the heavy neutrino $N_R$,
\\
(b)  There can also be light-heavy neutrino mixing 
diagrams with amplitudes $\sim m_D/M_R$ ,
\\
(c) Additional diagrams can also arise due to $W_L -W_R$ mixing,
\\
(d) Apart from above there can be additional contributions due 
to the  charged Higgs fields 
belonging to the bidoublet and triplet representations. \\
We discuss the relevant Lagrangian and the amplitudes of various 
contributions in the Appendix. 
In what follows we separate our discussion into three parts (i) type-I dominance, 
(ii) type-II dominance and (iii) contribution from the triplet Higgs. 

\subsection{\textcolor{green}{Type-I Dominance}}

In this section we discuss the $\nonubb$ for LR symmetric models
with type-I dominance with 
the $W_R$ mass to be of the order of 
$\sim$ TeV. 
The type-I term dominates if one takes $v_L$ = 0. It is well known that the
minimization of the potential in the LR symmetric model admits this possibility \cite{goran_vLzero}.
Then,
\begin{eqnarray}
m_{\nu}^{light} &=&   \frac{v^2}{v_R}y_D^T f^{-1}y_D ,\\
 m_{R}^{heavy}  &=&   f v_R ,
\end{eqnarray}
where we have used $f_L = f_R=f$.
If $U_R$ is the unitary matrix that diagonalizes the heavy neutrino mass
matrix $m_R$, then since $v_R$ is a constant, the same $U_R$ will also diagonalize $f$ and therefore one can write
\be
\Rightarrow f^{-1}= U_R^T(f^{dia})^{-1}U_R
\ee
Using this in the expression for $m_\nu^{light}$ one gets
\begin{eqnarray}
U_Rm_{\nu}U_R^T & = & \frac{v^2}{v_R}
U_Ry_D^T U_R^T(f^{dia})^{-1}U_Ry_D U_R^T 
\\ \nonumber
& = & \frac{v^2}{v_R}y_D^T (f^{dia})^{-1} y_D 
\\ \nonumber
& = & m_{\nu}^{dia}
\end{eqnarray}
Where in the last line we have used
$U_Ry_DU_R^T=y_D$ which is off course a special choice to establish a simplified relation between light and heavy neutrino masses.

The above choice implies
\be
U_L=U_R, ~~ \mathrm{and} ~~~
 m_{i}\propto\frac{1}{M_i}.\\
\label{typeI}
\ee
The proportionality constant depends on the vev's of Higgs which is generation independent and also $y_D$ which depends on $i$. 

In order to write the expression for the half-life of $\nonubb$ process
one needs to know the relative contribution of different terms. 
Since the mass of the right-handed gauge boson 
is related to the scale $v_R$, therefore for $M_{W_R} \sim$ TeV
the mass of the right-handed neutrinos is also $\sim$ TeV scale. 
However since $m_D^2/M_R \sim 0.01 - 0.1$ eV, $m_D^2$ should be 
$\sim 10^{10} - 10^{11}$ eV$^2$ which in turn implies $m_D/M_R 
\sim 10^{-5} - 10^{-6}$. Thus light-heavy mixing remains very small 
unless one does some fine tuning of the Yukawa textures \cite{tevseesaw}.
In this approximation, the only processes that contribute to the 
$\nonubb$ are the $W_L$ mediated diagrams through  light neutrino 
exchange and $W_R$ mediated diagrams through heavy neutrino exchange, as 
discussed in the Appendix.

The $\Delta_R$ mediated diagram 
can in principle contribute for $W_R$ mass around TeV scale.
But invoking the constraint from Lepton Flavour Violating(LFV) decays
it is seen that for majority of the parameter space $M_N/M_\Delta < 0.1$ and 
hence 
the $\Delta_R$ contribution can be suppressed \cite{Tello:2010am,petcovonnb}.
We will comment later on this scenario for the case $M_N \approx M_\Delta$. 

Under the above approximations the time-period for $\nonubb$ process
can be written as:
\begin{equation}
\frac{\Gamma_{  \nonubb }}{\text{ln\,2}} = {G} 
\Big(|\mathcal{M}_{\nu}|^2 |  
\eta_L |^2 +  
|\mathcal{M}_{N}|^2  |
\eta_R |^2\Big) ,  
\label{gamma} 
\end{equation} 
where 
\be
|\eta_L| = \frac{|U_{Lei}^2 m_i|}{m_e} = m_\nu^{ee}/m_e ,
\ee
\begin{equation} 
|\eta_R|  =\frac{M_{W_L}^4}{M_{W_R}^4} |(U_R)_{ei} ^2 m_p/M_i|.    
\end{equation}
The interference diagram between these two terms is helicity suppressed 
being proportional to the electron mass. 
The  contribution from the 
neutrino propagator term to the amplitude is 
  $\sim \frac{m_i}{p^2-m_i^2}$
The different dependence on the masses for the left and right 
sector come since the exchanged momentum $p$ satisfies 
$ m_i << p << M_i$.  
$\mathcal{M}_{\nu}$ and $\mathcal{M}_{N}$
are the nuclear matrix elements corresponding to light and heavy neutrino 
exchange respectively. 

Eq.(\ref{gamma}) can be  expressed as, 
\be
\frac{\Gamma_{\nonubb }}{\text{ln\,2}} = \frac{G 
|\mathcal{M}_{\nu}|^2 |}{m_e^2} |m_{ee}^{\mathrm{eff}}|^2.
\ee
Therefore the effective neutrino mass governing neutrinoless double beta decay is
\begin{eqnarray}
|m^{\rm eff}_{\rm ee}|^2
                      & = & |  m_{\nu}^{ee} |^2 + | M^{ee}_N |^2  , 
\label{eqn:Mnee_nldbd+lr}    
\end{eqnarray}
where, 
\be
| M^{ee}_N | = \bigg| <p>^2 \frac{M_{W_L}^4}{M_{W_R}^4}\frac{(U_R)_{ei} ^2}{ M_i } \bigg|,
\ee 
is the contribution to effective mass from the right-handed current.
In the above expression, 
$\big | <p^2> \big |= \big | m_e m_p{ \mathcal{M}}_{N}/{\mathcal{M}}_{\nu}\big |$. 

\begin{figure}[htb]
\begin{minipage}[t]{0.48\textwidth}
\hspace{-0.4cm}
\begin{center}
\includegraphics[width=5cm,angle=-90]{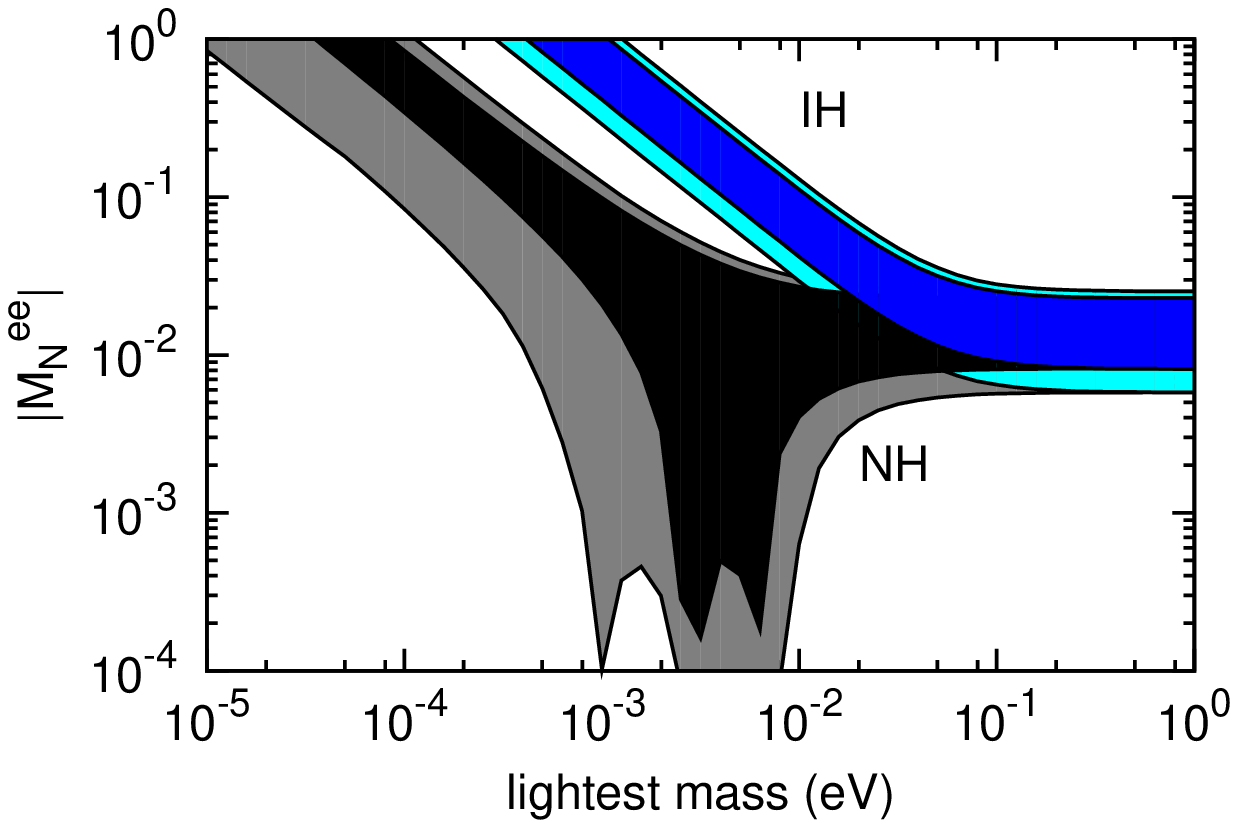}
\end{center}
 \end{minipage}
 \hfill
 \begin{minipage}[t]{0.48\textwidth}
 \hspace{-0.4cm}
 \begin{center}
 \includegraphics[width=5cm,angle=-90]{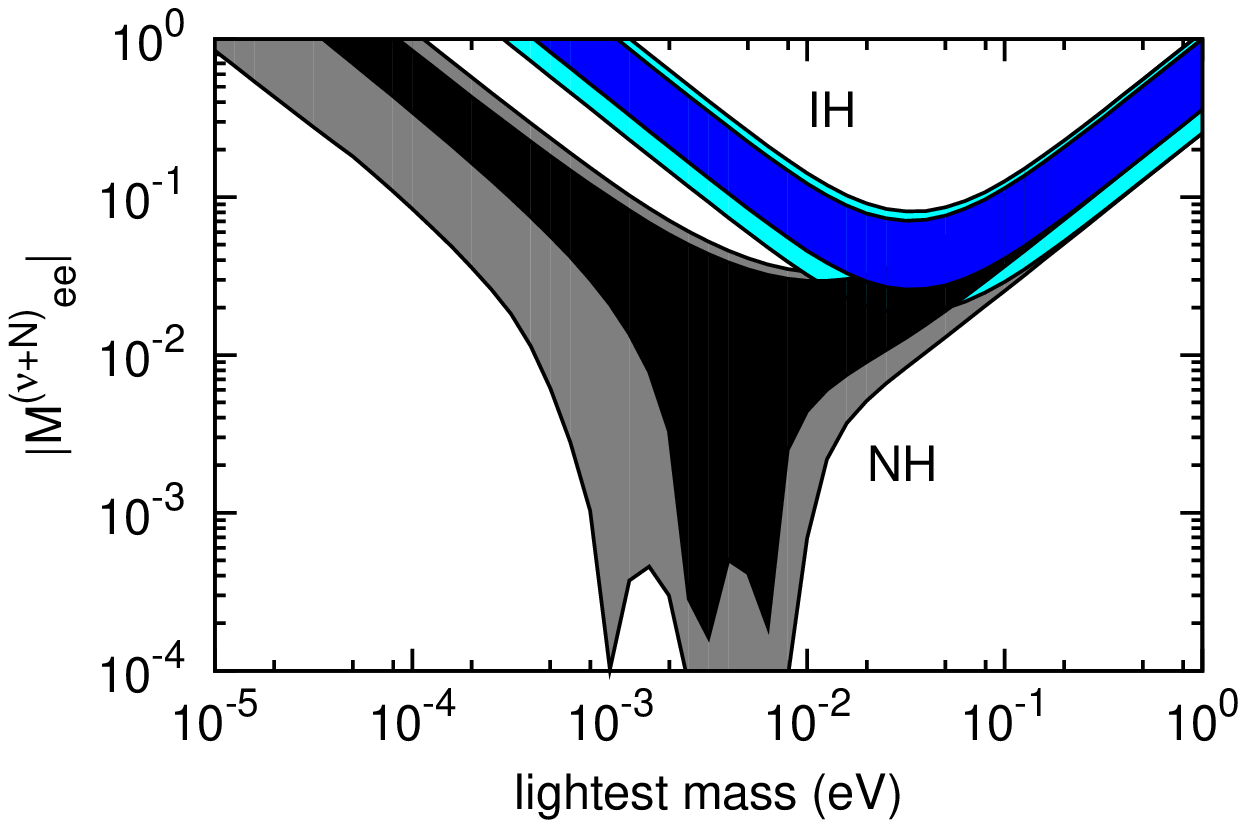}
 \end{center}
 \end{minipage}
 \caption{Contributions to effective mass $|M_{\scriptscriptstyle {N} }^{ee}|$ from only right-handed neutrino (left) and
the total contributions $|M_{\scriptscriptstyle {N+\nu} }^{ee}|$ from left+right handed neutrinos (right) in case of type-I
dominance.}
\label{typeI-plots}
 \end{figure}

In Fig.(\ref{typeI-plots}) we show the effective mass in LR symmetric 
models under type-I seesaw dominance. The right-handed neutrino 
mass $M_{W_R}$ is taken as 3.5 TeV. The mass of the heaviest right-handed
neutrino is 500 GeV. 
The allowed value of p is in the range $\sim $(100-200) MeV. In our analysis we have adopted 
p =180 MeV, $ M_{W_L}= 81$ GeV, 
$C_N = p^2(\frac{M_W}{M_{W_R}})^4 \sim 10^{10}~\mbox{eV}^2$.
The left panel shows the contribution from the right-handed current whereas 
the right panel shows the total contribution. For these plots we have used 
the 3$\sigma$ range of $\sin^2\theta_{13}$ as from the Daya-Bay results. 
Other oscillation parameters are varied in their 
$3\sigma$
ranges as given in Table 
\ref{table-osc}.
In order to understand the  various features of these diagrams and 
the interplay of the different contributions we examine the 
expression of effective mass in various limits for NH, IH and 
QD neutrinos. 

\subsubsection{\textcolor{blue}{Normal Hierarchy} }
In normal hierarchy regime as $m_1$ is the lightest neutrino, the heaviest 
RH neutrino will be $M_1$ as seen from the above Eq.(\ref{typeI}). 
The RH neutrino masses $M_2,M_3$ can therefore be expressed in terms of the heaviest RH mass as
\begin{eqnarray}
\frac{M_2}{M_1}=\frac{m_1}{m_2},\\
\frac{M_3}{M_1}=\frac{m_1}{ m_3}.
\end{eqnarray}
To find the above simplified relation between light and heavy neutrino masses we 
consider the eigenvalues of $y_D$ matrix to be degenerate, i.e., $y_{D1}=y_{D2}=y_{D3}$.
We get  the effective mass due to contribution from the RH neutrinos using above equation
Eq.(\ref{typeI})  
\begin{eqnarray}
\Big| M^{ee}_{N}\Big|
 &= & C_N \Big|\frac{ c_{12}^2c_{13}^2 }{M_1}+ \frac{s_{12}^2c_{13}^2}{M_2} e^{2i\alpha_2}+ \frac{s_{13}^2}{M_3} e^{2i\alpha_3}\Big|\\
&=& \frac{C_N}{M_1} \Big |  c_{12}^2c_{13}^2 
+ \frac{m_2}{m_1}s_{12}^2c_{13}^2e^{2i\alpha_2} 
+ \frac{m_3}{m_1}s_{13}^2e^{2i\alpha_3} \Big| .
\end{eqnarray}
In the limit when $m_1<<m_2 \simeq \sqrt{\dms} << m_3\simeq \sqrt{\dma}$ 
the above term reads as 
\be
\Big| M^{ee}_{N}\Big|=\frac{C_N}{M_1} \Big |  c_{12}^2c_{13}^2 
+ \frac{\sqrt{\dms}}{m_1}s_{12}^2c_{13}^2e^{2 i \alpha_2} 
+ \frac{\sqrt{\dma}}{m_1}s_{13}^2e^{2 i \alpha_3} \Big| .
\ee
This gives a steep increase in effective mass as we go towards the smaller
$m_1$. This is in complete contrast with the effective mass term 
$|m_{\nu}^{ee}|$. 
In this region therefore the total contribution is dominated by the RH sector.

As $m_1$ increases and to $m_1 \simeq m_2 \simeq \sqrt{\dms} << m_3 \simeq \sqrt{\dma}$. The effective mass term for the right-handed neutrino is
\begin{eqnarray}
\Big| M^{ee}_{N}\Big|
&=& \frac{C_N}{M_1} \Big |  c_{13}^2 ( c_{12}^2 + s_{12}^2e^{2i \alpha_2})   
+ \sqrt{\frac{1}{r}}s_{13}^2e^{2i \alpha_3}  \Big|,
\end{eqnarray}
for  $\alpha_2=0 , \alpha_3=\pi/2$ we get
\be
\sqrt{r} = \tan^2 \theta_{13} ,
\ee
for  $\alpha_2=\pi/2 , \alpha_3=\pi/2$
\be
\sqrt{r} cos2{\theta_{12}}= \tan^2 \theta_{13}.
\ee
The first condition cannot be satisfied by the present values of the oscillation parameters as is evident from Table \ref{table-cancellation}. But the 
second condition is same as what we have got for the light neutrino case and 
cancellations can occur in this range  for the right handed sector as well.

\subsubsection{\textcolor{blue}{Inverted Hierarchy} }
In inverted hierarchy regime  the lightest neutrino is the $m_3$. Therefore the heaviest RH neutrino in this ordering will be $M_3$ as is evident
from  Eq.(\ref{typeI}). 
The RH neutrino masses $M_1,M_2$ can therefore be expressed in terms of the heaviest RH Mass as 
\begin{eqnarray}
\frac{M_2}{M_3}=\frac{m_3}{m_2},\\
\frac{M_1}{M_3}=\frac{m_3}{ m_1}.
\end{eqnarray}
The heavy neutrino contribution to the effective mass is given by 
\be
\Big| M^{ee}_{N}\Big| = \frac{C_N}{M_3} \Big | \frac{m_1}{m_3} c_{12}^2c_{13}^2 
+ \frac{m_2}{m_3}s_{12}^2c_{13}^2e^{2i\alpha_2} 
+ s_{13}^2e^{2i\alpha_3} \Big| .
\ee
In the limit $m_3 << \sqrt{\dma} << m_2 \simeq m_1 \simeq \sqrt{\dma}$ the RH contribution to the effective mass is 
\be
\Big| M^{ee}_{N}\Big| = \frac{C_N}{M_3} \Big | \frac{\sqrt{\dma}}{m_3} c_{12}^2c_{13}^2 
+ \frac{\sqrt{\dma}}{m_3}s_{12}^2c_{13}^2e^{2i\alpha_2} 
+ s_{13}^2e^{2i\alpha_3} \Big| .
\ee
In the above equation the first and the second terms dominate and the 
effective mass reveal  an increase with decrease in $m_3$ for the IH case as well. 

\subsubsection{\textcolor{blue}{Quasi Degenerate} }

In this limit 
 $m_1 \simeq m_2\simeq m_3$ which in turn implies 
$M_1 \approx M_2 \approx M_3 \approx M_0$.  
The RH contribution to the effective mass can be expressed as,  
\be
\Big| M^{ee}_{N}\Big|=\frac{C_N}{M_0} \Big |  c_{12}^2c_{13}^2 + s_{12}^2c_{13}^2e^{2i \alpha_2}   + s_{13}^2e^{2i \alpha_3}  \Big|,
\ee
Thus the effective mass is independent of the light neutrino mass 
and remains constant. 
Since the light neutrino contribution in this limit increases with the 
mass scale this part dominates in the total contribution 
resulting in an overall increase in the total effective mass with increasing 
$m_1$ .

\noindent
\subsection{\textcolor{green}{Type-II dominance}}

Type-II dominance implies
the Dirac term connecting the left- and the
right- handed states is negligibly small as compared to the type-I term.
In this limit we can write
\begin{eqnarray}\nonumber
m_{\nu}^{light} &=&  f_Lv_L ,  \\\nonumber
 m_{R}^{heavy}  &=&  f_R v_R  .
\label{type2mass}
\end{eqnarray}
Denoting the matrices diagonalizing $m_{\nu}^{light}$ and $ m_{R}^{heavy}$
as $U_L$ and $U_R$ respectively, one can  have two possibilities
\begin{eqnarray} 
f_L  & = & f_R \Longrightarrow  U_L = U_R ,
\label{r1} 
\\
f_L  & = & f_R^{\ast} \Longrightarrow U_L = U_R^{\ast} .
\label{r2} 
\end{eqnarray}
Using Eq.(\ref{type2mass}) relation between
the light and heavy masses as,
\begin{equation} 
m_{i} \propto M_{i}.
\label{typeIImass} 
\end{equation}
Note that the proportionality constant in this case is $v_L/v_R$
and is independent of the generation index $i$.
In this case, using Eq.(\ref{typeIImass}) one can relate the heavy neutrino mass ratios to
those of light neutrinos as,  
\begin{eqnarray}
& & \frac{M_{_1}}{M_{_3}} = \frac{m_{1}}{m_{3}} \nonumber \\
&\text{and}& \quad \frac{M_{2}}{M_{3}} = \frac{m_{2}}{m_{3}}. \nonumber
\label{eqn:Mne} 
\end{eqnarray}
The dominant contributions to  $\nonubb$ come 
from two diagrams one via exchange of the light neutrinos
and another via exchange 
of the heavy neutrinos.   The charged Higgs diagram as well 
as the $W_L- W_R$ mixing diagrams can be neglected in this limit 
\cite{petcovonnb,Tello:2010am}. 
Therefore the time-period and the effective mass is given 
by the same  expression as in Eq. (\ref{eqn:Mnee_nldbd+lr}) 
for the type-I case.

\begin{figure}[htb]
\begin{minipage}[t]{0.48\textwidth}
\hspace{-0.4cm}
\begin{center}
\includegraphics[width=5cm,angle=-90]{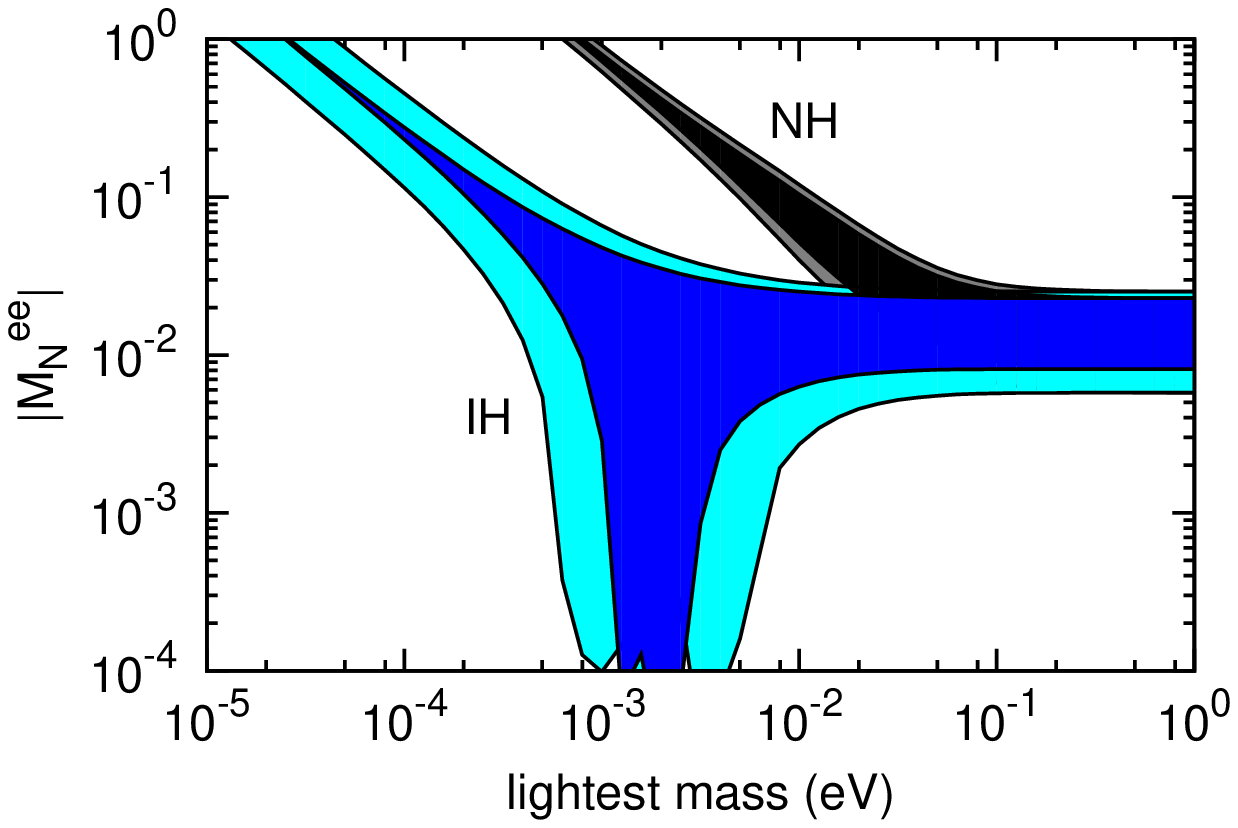}
\end{center}
 \end{minipage}
 \hfill
 \begin{minipage}[t]{0.48\textwidth}
 \hspace{-0.4cm}
 \begin{center}
 \includegraphics[width=5cm,angle=-90]{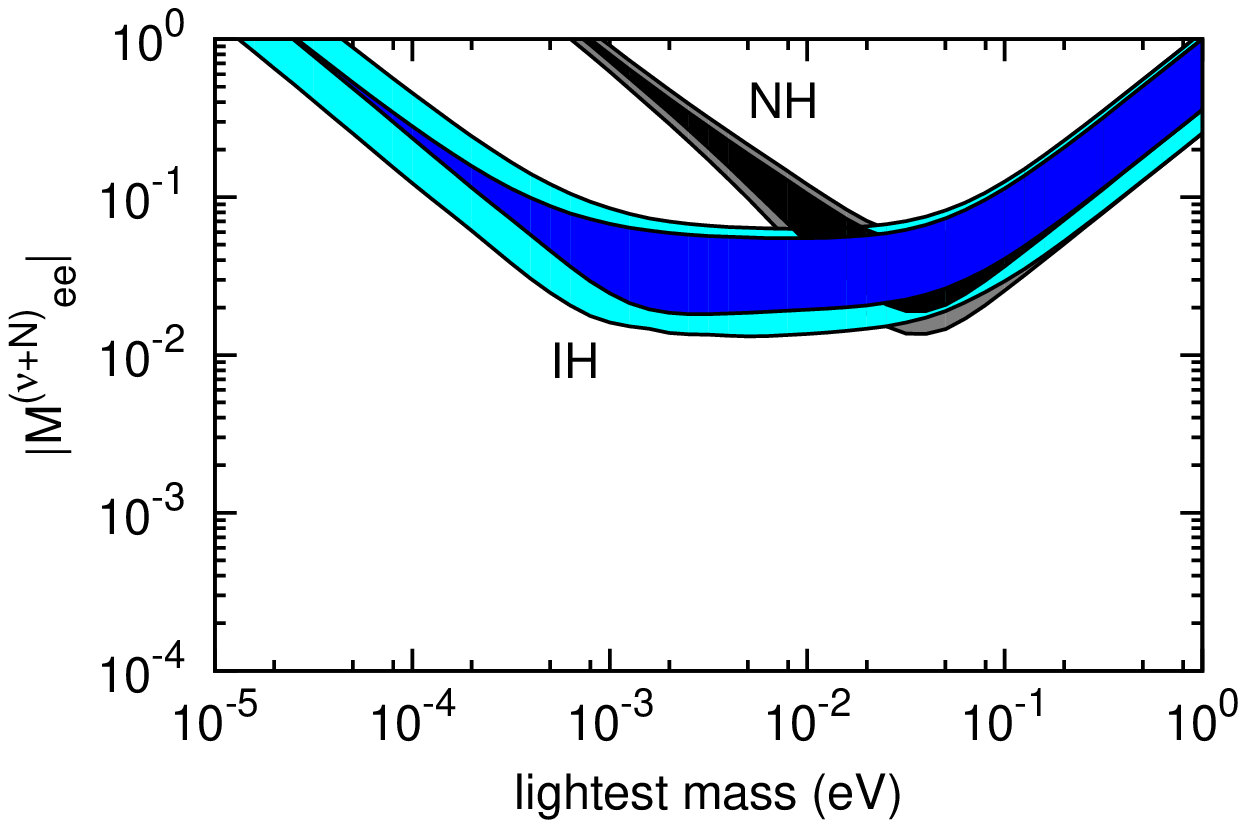}
 \end{center}
 \end{minipage}
 \caption{ Contributions to effective mass $|M_{\scriptscriptstyle {N} }^{ee}|$ from only right-handed neutrino (left) and
the total contributions $|M_{\scriptscriptstyle {N+\nu} }^{ee}|$ from left+right -handed neutrinos (right) in case of type-II
dominance.}
\label{typeII-plots}
 \end{figure}

Below we discuss the contributions to the effective mass from the 
heavy sector and the  interplay of the 
light and heavy contribution to the 
total effective mass for NH, IH and QD neutrinos.

\subsubsection{\textcolor{blue}{Normal Hierarchy}}

In order to examine the behavior of the  RH current, 
we fix $M_{3}$ (say around 500 GeV).
Once $M_{3}$ is fixed, the other heavy neutrino masses
can be expressed in terms of light neutrino data and
$M_{3}$.  Also, one can assume that $U_R =U_L$. 
We have assigned the same values of $p$, $M_{W_L}$ and $M_{W_R}$ 
as mentioned in type-I 
dominance case. The expression for $M^{ee}_N$ now becomes, 
\begin{eqnarray}
|M^{ee}_N|_{NH}  
&=& \frac{C_N}{M_{3}}  \bigg| c_{12}^2 c_{13}^2 \frac{m_3}{m_1} 
 + s_{12}^2 c_{13}^2 e^{2 i \alpha_2}\,\frac{m_3}{m_2}
           + s_{13}^2\,e^{2 i \alpha_3} \bigg|.
\label{eqn:M} 
\end{eqnarray}
In the limit of strict hierarchy:   
$ m_1 << m_2 \approx \sqrt{\Delta m^2_{\mathrm sol}} << m_3 \approx 
\sqrt{\Delta m^2_{atm}}$, 
the heavy neutrino contribution $M^{ee}_N$ can be written as, 
\begin{eqnarray}
|M^{ee}_N| = \frac{C_N}{M_{3}}  \bigg| c_{12}^2 c_{13}^2 \frac{ \sqrt{\Delta m^2_{\rm atm}} }{m_1}
+ s_{12}^2 c_{13}^2 \,e^{2 i \alpha_2} \frac{1}{\sqrt{r}} 
+  s_{13}^2\,e^{2 i \alpha_3} \bigg|.
\label{mNee_xnh} 
\end{eqnarray}
For smaller values of $m_1$, the  first term dominates showing 
a steep rise in the effective mass parameter as we go towards smaller 
$m_1$. This is shown in Fig.(\ref{typeII-plots}) in the left panel. 
This behaviour is in complete contrast with the effective mass 
$|m_\nu^{ee}|$ due to the  left-handed current which can
become vanishingly small due to complete cancellation between the various 
contributions.  
Since the heavy neutrino contribution is much higher, in the total effective 
mass, this plays the dominant role and hence the total effective mass shows 
a sharp decrease with $m_1$ in this regime as is seen from the right panel 
of Fig.(\ref{typeII-plots}). 
In the limit
$m_1 \approx m_2 \approx \sqrt{\Delta m^2_{sol}} << m_3 \approx \sqrt{\Delta m^2_{\rm atm}}$ 
\begin{eqnarray}
|M^{ee}_N|_{NH} = \frac{C_N}{M_{3}}  \bigg| 
\frac{c_{13}^2}{\sqrt{r}}
\bigg( c_{12}^2 
+ {s_{12}^2}  \,e^{2 i \alpha_2}\bigg) +  s_{13}^2\,e^{2 i \alpha_3} \bigg|.
\end{eqnarray}
The minimum value correspond to 
$\alpha_2 =0, \alpha_3 = \pi/2$ or 
$\alpha_2 = \alpha_3 = \pi/2$. 
Complete cancellation would require 
\be
\sqrt{r}~ \tan^2 \theta_{13} = c_{12} ^ 2 \pm {s_{12}^2} .
\ee 
The left-hand side is a product of two small numbers and 
for the current 3$\sigma$ range of parameters the above condition is 
not satisfied (cf. Table \ref{table-cancellation}) 
and hence complete cancellation cannot occur in this limit as 
well. As a result in the total effective mass the heavy neutrino contribution 
dominates in this regime.

\subsubsection{\textcolor{blue}{Inverted Hierarchy}}

In order to examine the analytical behavior for RH current for IH 
we fix the  highest mass state $M_{2}$  around 500 GeV.
Once $M_{2}$ is fixed, the other heavy neutrino masses
can be expressed in terms of light neutrino masses and
$M_{2}$ as
\begin{eqnarray}
& &\frac{ M_{1}}{ M_{2}} = \frac{m_{1}}{m_{2}} , \nonumber
             \\
&\text{and}& \quad \frac{M_{3}}{M_2} = \frac{m_{3}}{m_{2}}  .
              \nonumber 
\end{eqnarray}
Now the expression for $M^{ee}_N$ becomes
\begin{eqnarray}
|M^{ee}_N|_{IH} 
&=& \frac{C_N}{M_{2}}  \bigg| c_{12}^2 c_{13}^2 
\frac{m_2}{m_1} 
          + s_{12}^2 c_{13}^2 e^{2 i \alpha_2}
           + s_{13}^2\,e^{2 i \alpha_3} \,\frac{m_2 }{m_3}\bigg |.
\end{eqnarray}
For the smaller values of $m_3
<< \sqrt{\Delta m^2_{\rm atm}} \approx
m_2 \approx m_1 $, 
\begin{eqnarray}
|M^{ee}_N|_{IH} = \frac{C_N}{M_{2}}  \bigg| c_{12}^2 c_{13}^2 + s_{12}^2 c_{13}^2 \,e^{2 i \alpha_2}+ s_{13}^2
\frac{ \sqrt{\Delta m^2_{\rm atm}}}{m_3}
 \,e^{2 i \alpha_3}\bigg|.
\end{eqnarray}
It is clear from the above expression that for smaller values of $m_3$  
the absolute value of effective neutrino
mass varies with the lightest neutrino mass as $1/m_3$.
Since the contribution $m_\nu^{ee}$ in this region is smaller the 
total contribution is dominated by $|M^{ee}_N|$ and show the $1/m_3$ 
decrease. 
As $m_3$ increases the third term starts becoming smaller and there 
can be complete cancellations. 
There are
two end points corresponding to $\alpha_2=0, \pi, ~\alpha_3=\pi/2$ and $\alpha_2 =\pi/2, 
~\alpha_3=\pi/2$) where $M^{ee}_N$ acquires the minimum value.  This leads to 
the following  two conditions
for cancellation region
\begin{eqnarray}
& & m_3 =  \sqrt{\Delta m^2_{\rm atm}} \tan^2\theta_{13} 
\label{c1} 
 \\
& & m_3 = \sqrt{\Delta m^2_{\rm atm}} \tan^2\theta_{13} cos2\theta_{12}.
\label{c2} 
\end{eqnarray}
The Eq.(\ref{c1}) gives $m_3 \sim 6 \times 10^{-5} - 2 \times 10^{-3}$ eV
for the 3$\sigma$ ranges of parameters from Table \ref{table-osc} (but $\sin^2 \theta_{13}$ from Daya-Bay)
while  
Eq.(\ref{c2}) gives $m_3 = 1.7 \times 10^{-4} - 9.2 \times 10^{-4} $.  
These ranges define the values of $m_3$ for which complete cancellation 
can occur. This is reflected in Fig(\ref{typeII-plots}). 
However when we consider the total effective mass then the contribution from 
$|m_\nu^{ee}|$ dominates and  enforces a lower limit on this. 
Thus vanishing effective mass is not a possibility for both hierarchies
in presence of right-handed currents and type-II seesaw dominance. 

\subsubsection{\textcolor{blue}{Quasi Degenerate}} 

Quasi-degenracy in the light neutrino masses also 
implies   
$M_1 \approx M_2 \approx M_3 \approx M_0$ 
for the heavy neutrinos. 
In this regime 
the heavy neutrino contribution to the effective mass is 
\begin{eqnarray}
|M^{ee}_N|_{QD} = \frac{C_N}{M_{0}}  \bigg|c_{12}^2 c_{13}^2 + s_{12}^2  c_{13}^2
\, e^{2 i \alpha_2} + s_{13}^2 \,e^{2 i \alpha_3} \bigg|.
\end{eqnarray}
This is independent of the lightest neutrino mass
and hence value of $|M^{ee}_N|$ remains constant with increasing 
value of the common mass scale.
The overall behaviour of the total effective mass in this regime 
is therefore controlled by the lighter neutrino contribution and increases 
with increasing mass. 

\subsection{\textcolor{green}{The contribution from the triplet Higgs}}

\begin{figure}[htb]
\begin{minipage}[t]{0.48\textwidth}
\hspace{-0.4cm}
\begin{center}
\includegraphics[width=5cm,angle=-90]{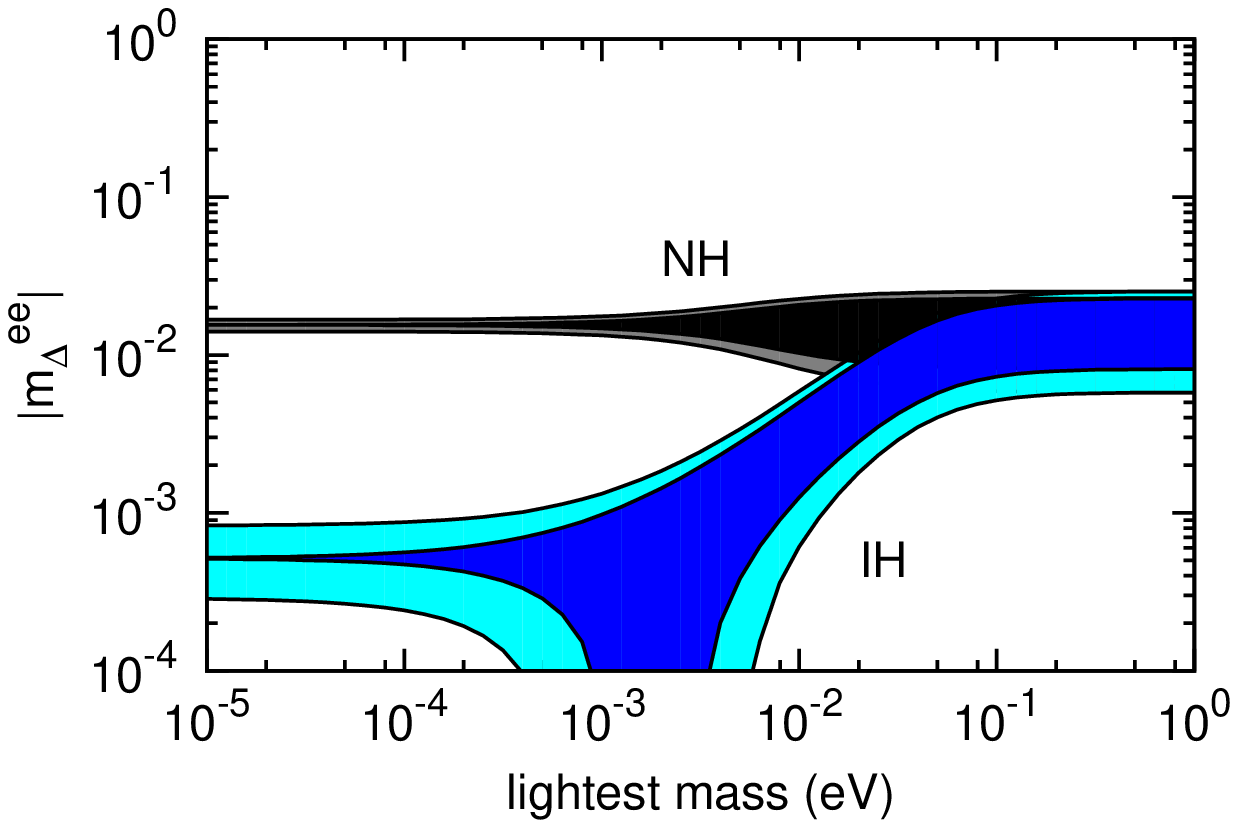}
\label{fig:4a}
\end{center}
 \end{minipage}
 \hfill
 \begin{minipage}[t]{0.48\textwidth}
 \hspace{-0.4cm}
 \begin{center}
 \includegraphics[width=5cm,angle=-90]{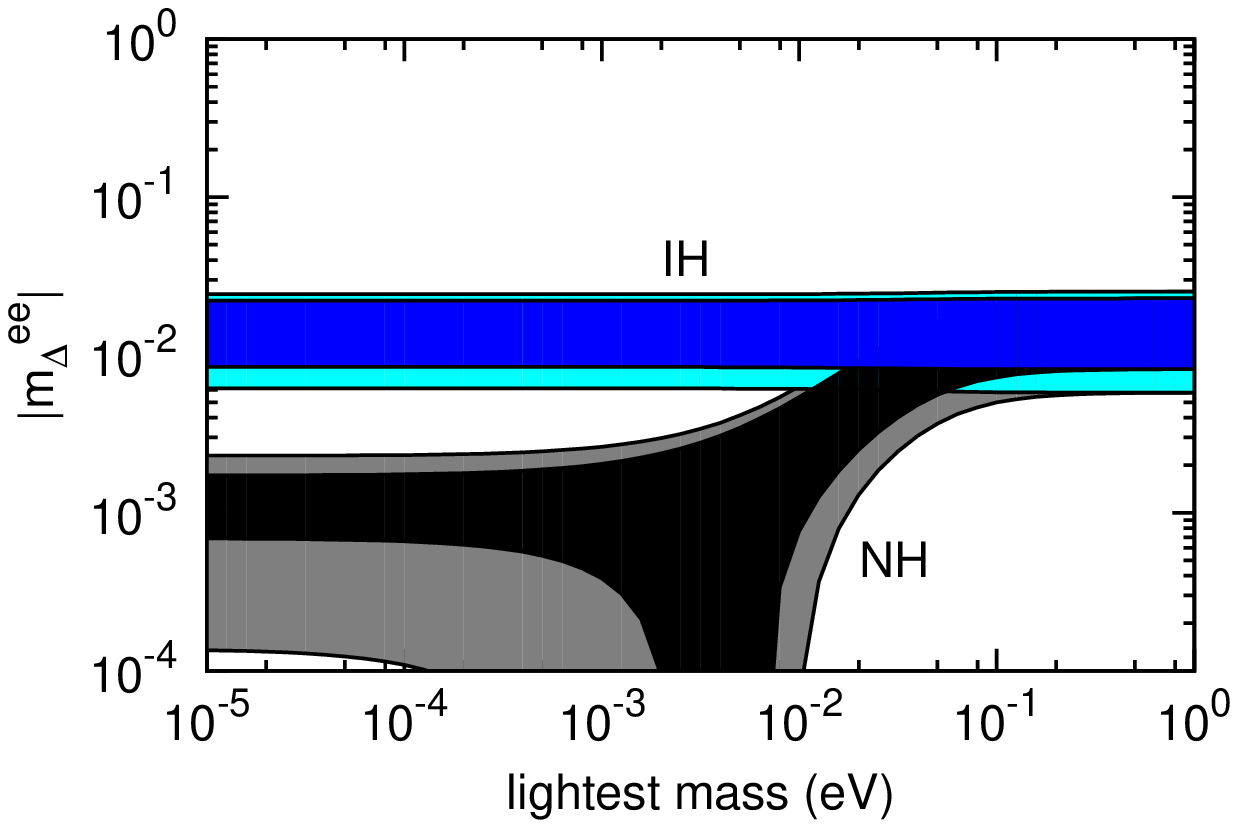}
 \label{fig:4b}
 \end{center}
 \end{minipage}
 \caption{The contribution to the effective mass from 
right-handed triplet diagrams for type-I seesaw (left) and type-II 
seesaw (right) diagrams respectively.}
\label{typeII-lr-2charged-scalar}
 \end{figure}

The Majorana masses of light and heavy neutrinos comes naturally in left-right model because
of the two triplets $\Delta_{L,R}$. 
As discussed in the appendix the contribution from $\Delta_L$ is 
much suppressed as compared to the dominant contributions. 
However the magnitude of the $\Delta_R$ contribution is controlled by the 
factor  $M_i/M_{\Delta_{R}}$. 
In the earlier sections we have not included the contribution due to
the triplet Higgs diagrams under the assumption $M_i/M_{\Delta_{R}} <0.1$ 
, which is obtained from LFV processes \cite{Tello:2010am}. However,
this approximation though valid in a large part of the parameter space
there are some allowed mixing parameters for which this ratio can be 
higher \cite{Tello:2010am} .
 In that case the contribution from this diagram needs 
to be included. In order to asses the impact of this contribution 
below we discuss the effective mass due to the $\Delta_R^{--}$ diagram 
(see Appendix) in the limit $M_{heaviest} = M_{\Delta_{R}}$. 

Comparing with the light neutrino exchange diagram,
the $\Delta_R$  exchange
diagram (which is shown in Fig.(\ref{typeII-lr-2charged-scalar})) gives
the effective mass as, 
\begin{eqnarray}
&|m^{ee}_\Delta|&=  \bigg|p^2\, \frac{M^4_{W_L}}{M^4_{W_R}}\,\frac{2\, M_N}{M^2_{\Delta_{R}}}\bigg|.
\end{eqnarray}

In Fig.(\ref{typeII-lr-2charged-scalar}) we plot the contribution of the effective mass 
due to the triplets, $|m^{ee}_\Delta|$, for type-I and type-II seesaw model.
The figure shows that for the type-I case and NH the triplet contribution 
can dominate over the light and heavy neutrino contributions in the 
cancellation regime. However for the other cases, i.e., type-I and IH  and 
type-II (both NH and IH), 
the triplet contributions are always lower than the other two 
contributions. 

Note that in generating the above plots we have varied the mixing parameters
in their 3$\sigma$ range and all phases between $0 ~\mathrm{to} ~2\pi$.  
However, note that the approximation $M_{heaviest} = M_{\Delta_{R}}$ 
may not be valid for all the parameter 
values and a  more accurate analysis would require a correlated study 
of LFV and $\nonubb$ which is the beyond the scope of this paper.  

\section{Conclusions}
Neutrino oscillation experiments have already provided us the first signature 
of physics beyond the Standard Model.  At the present juncture the quest for
new physics has also got an unprecedented momentum because of LHC. 
A natural question is whether the origin of neutrino mass 
can be probed at LHC and  it is hoped that 
the mist around the TeV
scale may  be uplifted by  complimentary information from 
neutrino and collider experiments.   
Among neutrino experiments, observation of 
$\nonubb$  would signify lepton number violation and Majorana 
nature of neutrino mass.  However as is well known $\nonubb$ can also occur 
in many other scenarios and hence the specific nature of new physics 
may remain to be ascertained and in such situations LHC and LFV processes
may provide complimentary information. This interrelation of $\nonubb$ with 
LHC and LFV processes makes it a very engrossing and interesting topic 
of research at the present juncture. 
In this paper we  scrutinize the implications of 
$\nonubb$ in 
left-right symmetric models 
with right handed gauge bosons  and neutrinos around TeV scale.
We analyze the effective mass 
within the frameworks of type-I and type-II
dominance cases. 
For the latter case the hierarchy in heavy sector is same as that in 
the light sector. For the former, we use some simplifying assumptions for 
relating the mass hierarchy of the heavy sector with that of the light sector. 
We identify the situations where the dominant 
contribution can come from the right-handed sectors. 
The various parameters chosen for our numerical work
are consistent with constraints from neutrino oscillation experiments
as well as results from LHC  and Lepton Flavour Violation. 
In the light of recent announcement on the measurement of the third leptonic
mixing angle from reactor experiments
Daya Bay and RENO we have  discussed the impacts of 
the measured value of $\theta_{13}$ on the behaviour of effective mass. 
The regime of cancellations between different contributions to $\nonubb$ and their compatibility with 
present oscillation data are also presented. 
We note that in both type-I, and type-II dominance situations the right-handed
contribution can override the left-handed ones for smaller 
values of the lightest neutrino mass while in the quasi-degenerate regime the 
light neutrino contribution dominates. 
For the type-II dominance case 
the predictions for NH and IH 
including right-handed currents are in stark contrast  compared to the case 
of only left-handed currents. 
Thus, signature of TeV scale LR symmetry at LHC may present a
completely altered
interpretation for  effective mass governing $\nonubb$ process.

\section{Appendix} 

The canonical type-I seesaw in left-right symmetric model is automated through
the presence of the bidoublet that generates the Dirac coupling, $m_D$
between left- and right- handed neutrinos 
 while the type-II seesaw 
term $m_L$ is due to the presence of the triplet Higgs. 
The relevant Lagrangian reads as
\begin{equation}
\mathcal{L} = -\frac{1}{2}\, \left(\begin{array}{cc} \overline{\nu^{c\prime}_L} & \overline{N^\prime_R}\end{array} \right)\, 
\mathcal{M}
\left(\begin{array}{c}\nu^\prime_L \\
N^{c\prime}_R \end{array} \right) + \text{h.c},
\label{eq:seesaw1} 
\end{equation}
where $\mathcal{M}$ is the neutrino mass matrix which can be expressed as, 
\begin{equation}
\mathcal{M}=\left(\begin{array}{cc} 
m_L & m^T_D \\ 
m_D & M_R 
\end{array} \right)_{6 \times 6} .
\label{eq:seesawfull}
\end{equation}
$M_R$ is the Majorana mass matrix for the right-handed neutrinos which 
arises through the right handed triplet Higgs. 
The $6\times6$ neutrino mass matrix Eq.(\ref{eq:seesawfull}) can be diagonalized by the 
$6\times6$ unitary matrix $\mathcal{U}$, defined as
\begin{eqnarray}
\mathcal{U}^T\, \left(\begin{array}{cc} 
0 & m^T_D \\ 
m_D & M_R 
\end{array} \right) \mathcal{U} = \left(\begin{array}{cc} 
m^{diag} &  0 \\ 
0 & M^{diag} 
\end{array} \right)
\quad \text{and} \quad
\left(\begin{array}{c}
\nu^\prime_L \\
 N^{c\prime}_R \end{array} \right) =\mathcal{U} 
\left(\begin{array}{c}\nu_L \\ N^c_R \end{array}\right),
\end{eqnarray}
where $m^{diag}=\mbox{diag}(m_1,m_2,m_3)$ and $M^{diag}=\mbox{diag}(M_1,M_2,M_3)$ are the diagonal matrices
with mass eigenvalues $m_i$ and $M_i$ ($i=1,2,3$) for light and heavy neutrinos respectively.
The primed and unprimed notations in the neutrino fields correspond
to the flavour and mass eigenstates respectively. 
The complete unitary mixing matrix $\mathcal{U}$ can
be parametrized as a product of two matrices
$\mathcal{U}=W\, U_{\nu}$ and can be expressed as \cite{grimus} 
\begin{equation}
\mathcal{U} = W\, U_{\nu} 
            = \left(\begin{array}{cc} 
          (1-\frac{1}{2}RR^{\dag})\, U^\prime_L & R\, U^\prime_R\\
        -R^{\dag}\, U^\prime_L & (1-\frac{1}{2}\, R^{\dag}R)\, U^\prime_R
        \end{array} \right) = \left(\begin{array}{cc}
U_L & T\\
S & U_R
 \end{array} \right) ,
\label{u-unitary} 
\end{equation}
where $W$ is the matrix which brings the $6 \times 6$ neutrino matrix, Eq.(\ref{eq:seesawfull}),
in the block diagonal form
and 
$ U_\nu = diag(U^\prime_L, U^\prime_R)$
diagonalizes the left and right handed parts. 
The matrix $R$ appearing in the Eq.(\ref{u-unitary}) 
is $R = m^\dagger_D\, M^{-1 *}_R$.

In left-right symmetric theories, 
the charged current interactions of leptons, in the flavour basis
is given by
\begin{eqnarray}
\mathcal{L}_{\rm CC} &=& \frac{g}{\sqrt{2}}\, \sum_{\alpha=e, \mu, \tau}
\bigg[ \overline{\ell}_{\alpha \,L}\, \gamma_\mu {\nu^\prime}_{\alpha \,L}\, W^{\mu}_L 
      + \overline{\ell}_{\alpha \,R}\, \gamma_\mu {N^\prime}_{\alpha \,R}\, W^{\mu}_R \bigg] + \text{h.c.} 
\end{eqnarray}
Assuming  
a basis such that the charged-lepton mass matrix 
is diagonal, the above  can be
rewritten in the mass basis as, 
\begin{eqnarray}
\mathcal{L}_{\rm CC} &=& \frac{g}{\sqrt{2}}\, \sum_{\alpha=e, \mu, \tau}\, \sum_{i=1}^{3}
\bigg[ \overline{\ell}_{\alpha \,L}\, \gamma_\mu\, 
                    \{(U_L)_{\alpha i} \nu_{L i}+(T)_{\alpha i} N_{R i}^c\} W^{\mu}_L \nonumber \\
&\hspace{2cm}&
 + \overline{\ell}_{\alpha \,R}\, \gamma_\mu\, 
                     \{(S)^{*}_{\alpha i} \nu_{Li}^c+(U_R)^{*}_{\alpha i} N_{Ri}\} W^{\mu}_R
\bigg] + \text{h.c.}
\end{eqnarray}

\begin{figure}[htb]
 \label{fig:typeI-lr-WL}
 \includegraphics[width=15cm,height=5cm]{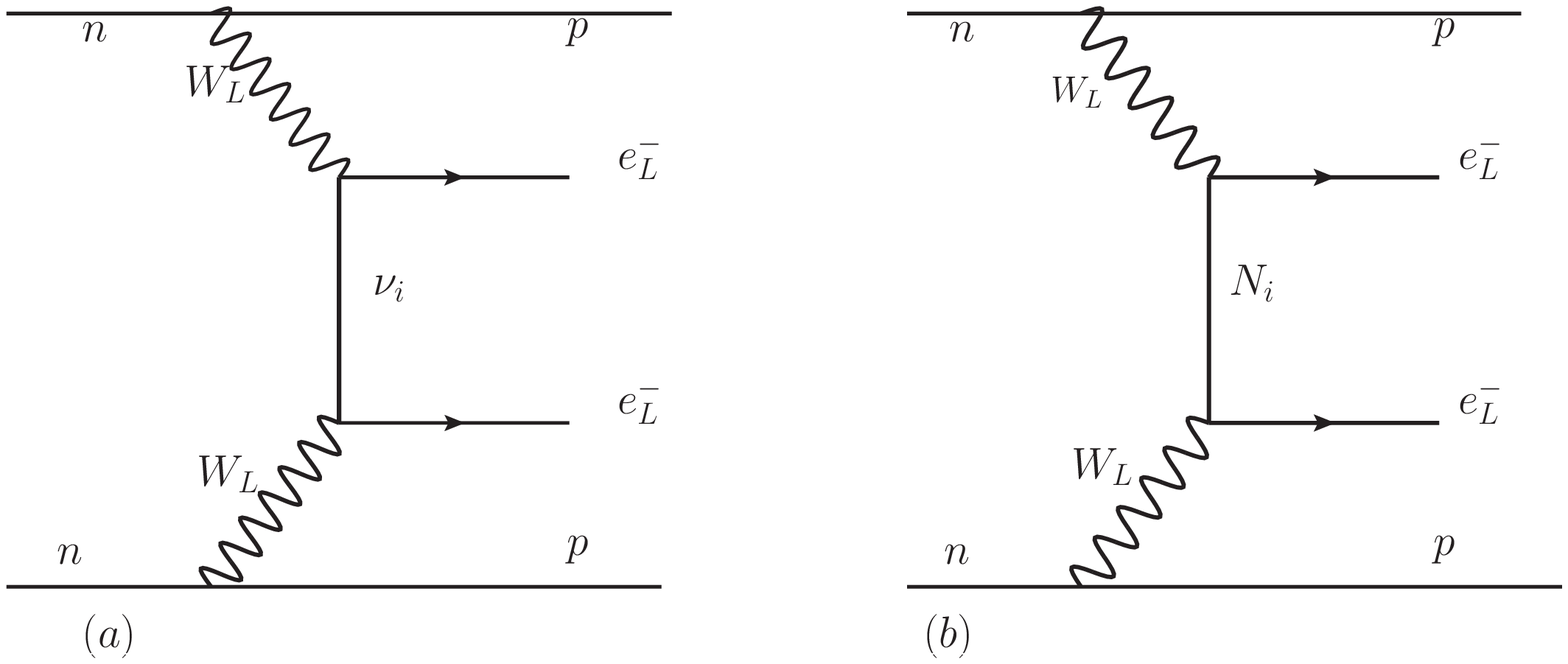}
 \caption{Neutrinoless double beta decay contribution from light and heavy
 Majorana neutrino intermediate states from two $W_L$ exchange. 
}

 \end{figure}
 \begin{figure}[htb]
 \label{fig:typeI-lr-WR}
 \includegraphics[width=15cm,height=5cm]{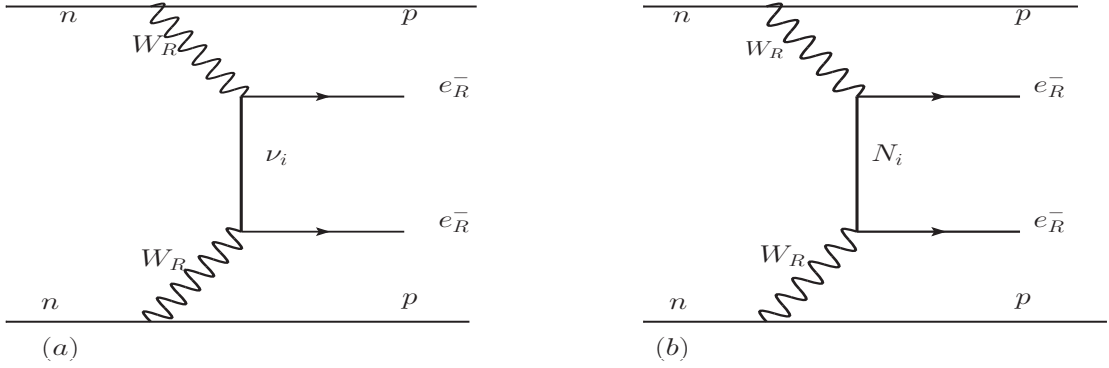}

 \caption{Neutrinoless double beta decay contribution from light and heavy
 Majorana neutrinos from two $W_R$ exchange. 
}

 \end{figure}
 \begin{figure}[htb]
 \label{fig:typeI-lr-mix}
 \includegraphics[width=15cm,height=5cm]{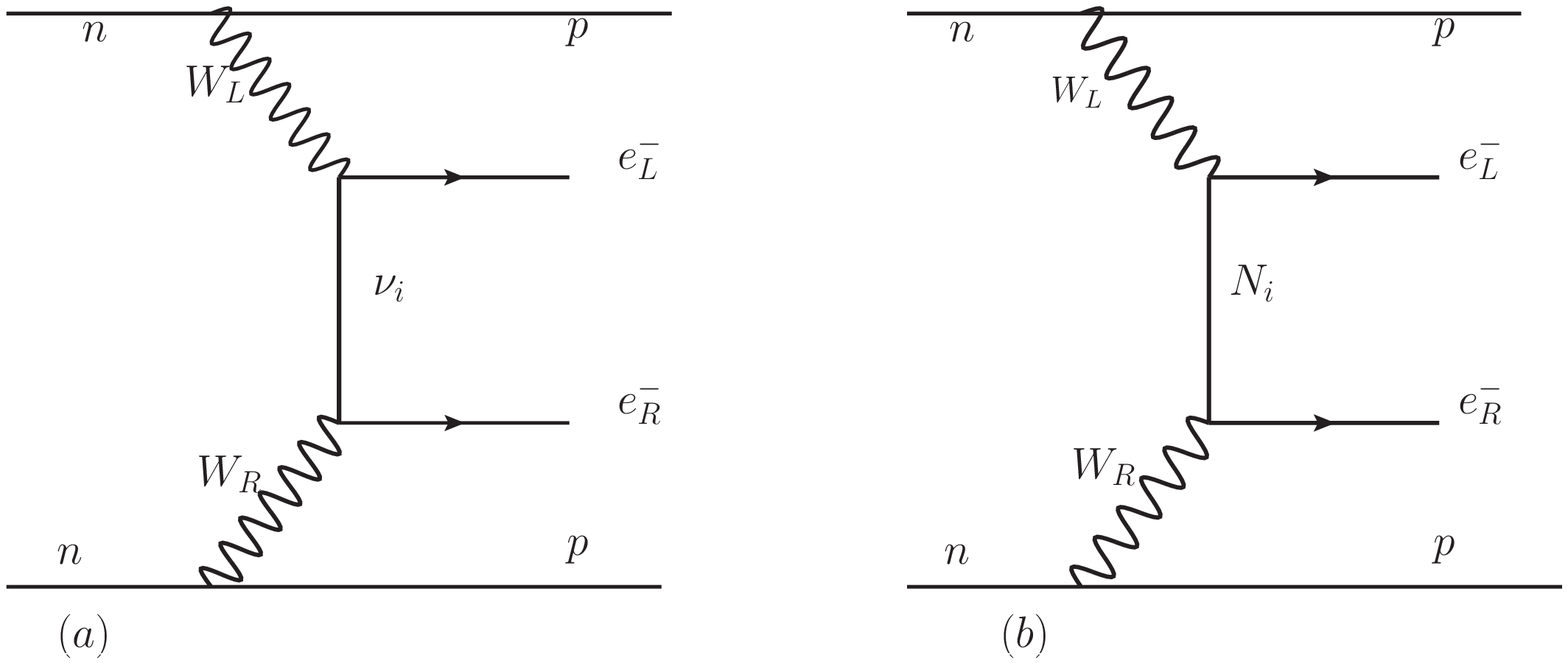}
 \caption{Neutrinoless double beta decay contribution from light and heavy
 Majorana neutrino intermediate states from both left- and right- handed gauge bosons
 exchange at each vertices's.}
\end{figure}

With this particular form of the charged current interaction, there are several additional diagrams
to neutrinoless double beta decay. We have categorised them as: first that involves two $W_L$ (the
amplitude is denoted as $\mathcal{A}^{LL}$), the second that involves the exchange of two $W_R$
gauge bosons $\mathcal{A}^{RR}$), and the third one that involves both $W_L$ and $W_R$ exchange at the
same diagram $\mathcal{A}^{LR}$).
The amplitudes of the contributions  from the different diagrams 
are as indicated below :  
\begin{enumerate} 
\item 
[(i)] ~~$\mathcal{A}^{LL}_{\nu} \propto \frac{1}{M^4_{W_L}} \frac{U^2_{L_{e\,i}}\, m_i}{p^2} ~~~~{\mathrm{(Fig. 6a)}}$, 

\item
[(ii)] ~~ $\mathcal{A}^{LL}_{N} \propto \frac{1}{M^4_{W_L}} \frac{T^2_{e\,i}}{M_i} $~~~ {\rm(Fig. 6b)},

\item
[(iii)]~~
${ \mathcal{A}^{RR}_{\nu} \propto \frac{1}{M^4_{W_R}} \frac{(S^*_{e\,i})^2\, m_i}{p^2}} $ ~~~
{\rm (Fig. 7a)} ,

\item
[(iv)]~~
${ \mathcal{A}^{RR}_{N} \propto \frac{1}{M^4_{W_R}} \frac{(U^*_{R_{e\,i}})^2}{M_i} }$
~~~{\rm (Fig. 7b)} ,

\item 
[(v)]~~
${ \mathcal{A}^{LR}_{\nu} \propto \frac{1}{M^2_{W_L}\,M^2_{W_R}} \frac{U_{L_{e\,i}} S^*_{e\,i}\, m_i}{p^2} }$
~~~{\rm (Fig. 8a)} ,

\item 
[(vi)] 
${ \mathcal{A}^{LR}_{N} \propto \frac{1}{M^2_{W_L}\,M^2_{W_R}} \frac{T_{e\,i}\, U^*_{R_{e\,i}}\,}{M_i}} $
~~~{\rm (Fig. 8b)} .

\end{enumerate}

There will also be the  contribution from the triplet Higgs diagrams
as given in Fig.(8).

\begin{figure}[htb]
 \label{fig:triplet-lr-WL}
 \includegraphics[width=15cm,height=5cm]{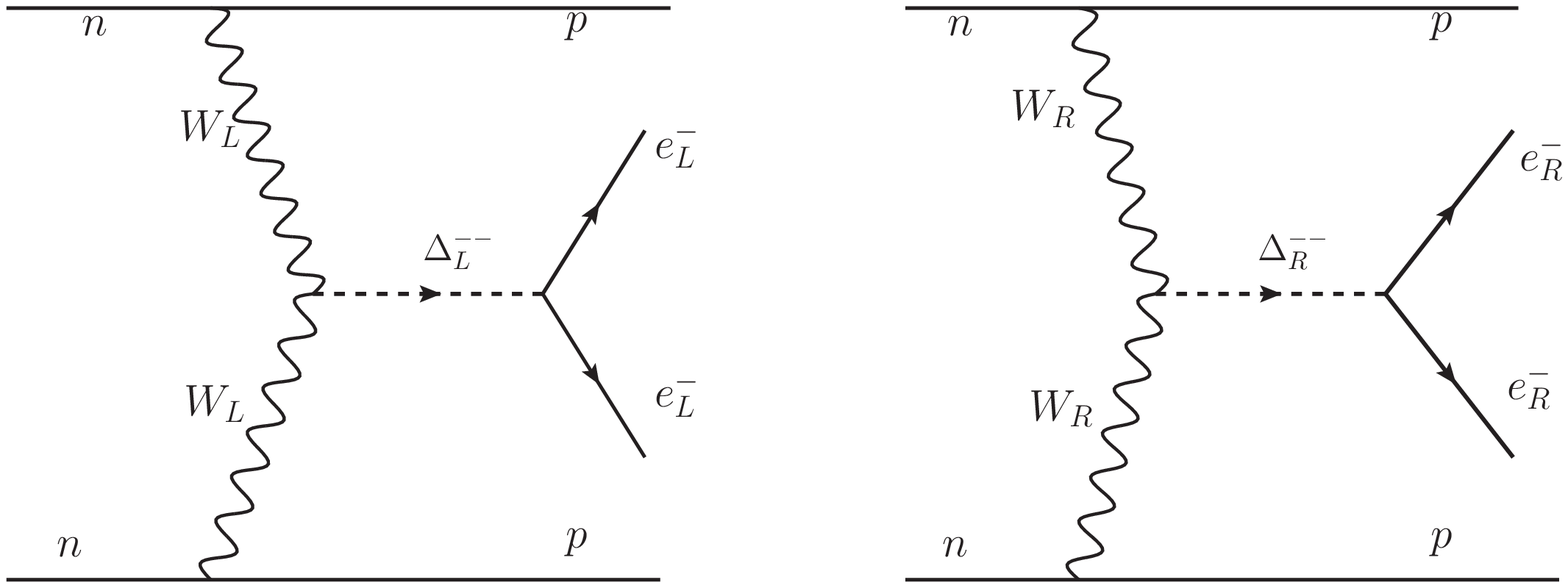}
 \caption{Neutrinoless double beta decay contribution from the charged Higgs
 intermediate states from $W_L$ and $W_R$ exchange.}
 \end{figure}

The amplitudes for the diagrams can be written as, 
\begin{enumerate} 
\item 
[(i)] ~~$\mathcal{A}^{LL}_{\Delta_L} \propto \frac{1}{M^4_{W_L}} 
\frac{1}{M_{\Delta_L}^2} f_L v_L $ ,

\item
[(ii)]~~
$ \mathcal{A}^{RR}_{\Delta_R} \propto \frac{1}{M^4_{W_R}}
\frac{1}{M_{\Delta_{R}^2}} f_R v_R $.
\end{enumerate}

The time period for $\nonubb$ can be expressed as, 
\begin{equation}
\frac{\Gamma_{  \nonubb }}{\text{ln\,2}} = {G} 
\Big|
\mathcal{M}_{\nu_L}\Big( \eta^{LL}_\nu  + \eta^{RR}_\nu + \eta^{LR}_\nu 
+\eta^{LL}_{\Delta_L}\Big)    
+ ~\mathcal{M}_{N_R} \Big(\eta^{LL}_N   +   
\eta^{RR}_N   +   
\eta^{LR}_N    
+ \eta^{RR}_{\Delta_R}\Big)
\Big |^2    ,
\label{tp} 
\end{equation} 

where the dimensionless parameters $\eta$ are defined as follows: 
\\
$$
\mathcal{\eta}^{LL}_{\nu} =  \frac{U^2_{L_{e\,i}}\, m_i}{m_e}; \nonumber \quad 
\mathcal{\eta}^{LL}_{N} =  \frac{T^2_{e\,i} m_p}{M_i} ;\nonumber 
$$
\\
$$\mathcal{\eta}^{RR}_{N} = \frac{M^4_{W_L}}{M^4_{W_R}} \frac{U^{*2}_{R_{e\,i}} m_p}{M_i}; \nonumber
\quad
\mathcal{\eta}^{RR}_{\nu} = \frac{M^4_{W_L}}{M^4_{W_R}} \frac{S^{*2}_{e\,i}\, m_i}{m_e} ;
$$\\
$$
\mathcal{\eta}^{LR}_{\nu} = \frac{M^2_{W_L}}{M^2_{W_R}} \frac{U_{L_{e\,i}} S^{*}_{e\,i}\, m_i}{m_e}; \nonumber
\quad
\mathcal{\eta}^{LR}_{N} = \frac{M^2_{W_L}}{M^2_{W_R}} \frac{T_{e\,i}\, {U^*}_{R_{e\,i}}\, m_p }{M_i}; \nonumber
$$
\\
$$\eta^{LL}_{\Delta_L} = \frac{{U^{2}_{Lei}} m_i m_e}{M_{\Delta_L}^2};
~~
\eta^{RR}_{\Delta_R} = \frac{M^4_{W_L}}{M^4_{W_R}} \frac{{U^{2}_{Rei}}^2 M_i m_p}{M_{\Delta_R}^2}.
$$
Inserting these the half-life is, 
\bea
\frac{\Gamma_{  \nonubb }}{\text{ln\,2}}  & =  & {G} 
\frac{|\mathcal{M}_{\nu }|^2}{m_e^2}  
\bigg | \Big(U^2_{L_{e\,i}}\, m_i +  
p^2 \frac{T^2_{e\,i}}{M_i} 
+ p^2 \frac{M^4_{W_L}}{M^4_{W_R}} \frac{U^{*2}_{R_{e\,i}}}{M_i}
\\ \nonumber 
&  &+ \frac{M^4_{W_L}}{M^4_{W_R}} {S^{*2}_{e\,i}\, m_i}
+  \frac{M^2_{W_L}}{M^2_{W_R}} U_{L_{e\,i}} S^*_{e\,i} m_i
+  p^2 \frac{M^2_{W_L}}{M^2_{W_R}}   \frac{T_{e\,i}\, U^*_{R_{e\,i}}\,}{ M_i}
\\ \nonumber
& & + \frac{{U^{2}_{Lei}} m_i m_e^2}{M_{\Delta_L}^2} 
+ p^2 \frac{M^4_{W_L}}{M^4_{W_R}} \frac{{U^{2}_{Rei}} M_i}{M_{\Delta_R}^2}
\Big) \bigg |^2.
\eea
In order to estimate the relative contributions of different terms note that 
our analysis is done  $M_{W_R} \sim$ TeV and the  heaviest 
right-handed neutrino  
is in the range $\sim$ 500 GeV. Since $m_\nu \sim  
m_D^2/M_R \sim 0.01 - 0.1$ eV, $m_D \approx  
10^{5}$ eV which in turn implies $m_D/M_R 
\sim 10^{-6} - 10^{-7} \sim T_{ei} \sim S_{ei}$ 
and $p^2 \sim 100$ MeV. We also assume as illustrative values 
$M_{\Delta_L} \approx M_{\Delta_{R}}$ = 1 TeV. 

Then the order of magnitude of the various terms in the above expression 
are as follows, 
\begin{itemize} 
\item $U_{ei}^2 m_i \sim m_i \sim 0.01 - 0.1 $ 
\item  $p^2 \frac{T^2_{e\,i}}{M_i} \sim 10^{-8}$ 
\item  
$p^2 \frac{M^4_{W_L}}{M^4_{W_R}} \frac{U^2_{L_{e\,i}}}{M_i} \sim 0.01$
\item 
$\frac{M^4_{W_L}}{M^4_{W_R}} {S^2_{e\,i}\, m_i} \sim 10^{-18} m_i$ 
\item 
$\frac{M^2_{W_L}}{M^2_{W_R}} {S_{e\,i}\, U_{R_{e\,i}}\, m_i} 
\sim 10^{-9} m_i $ 
\item 
$p^2 \frac{M^2_{W_L}}{M^2_{W_R}} \frac{T_{e\,i}\, U_{R_{e\,i}}}{ M_i} \sim 10^{-5} $
\item $\frac{U^{2}_{Lei} m_i m_e^2}{{M_{\Delta_L}}^2} \sim 10^{-13} m_i$
\item $p^2 \frac{M^4_{W_L}}{M^4_{W_R}} \frac{{U^{2}_{Rei}} M_i}{M_{\Delta_R}^2}
\sim  10^{-5}$ .
\end{itemize} 

Therefore, in the approximation the non-unitarity of the mixing is small 
($\sim 10^{-6}$) the dominant contribution will come from the left-handed 
current with light neutrino exchange and the right-handed current with the heavy neutrino exchange. 
However in cases where these contributions vanish due to 
some cancellations the effective mass can be generated  
from the diagram 8(b)  since the same cancellation condition may not 
be operative in this term \cite{xingnu}.    
However because of $T_{ei}$ term the overall contribution  
is still expected to be small for our choice of parameters.
In our analysis we will therefore 
neglect this contribution.  
Also, if one assumes $M_i \approx M_{\Delta_R}$ then 
the $\Delta_R$ exchange diagram can also become significant. 
We have made some comments on this situation in the main text. 
We have neglected $W_L- W_R$ mixing which is 
$ \leq {\mathcal{O}}(10^{-3})$ and would cause a further suppression. 

\section{Acknowledgment } 

S.G. would like to thank James Barry, Subrata Khan, 
Manimala Mitra and Werner Rodejohann for helpful discussions. 
S. Patra would like to thank the hospitality at PRL, where most of the present
work was done.

\end{document}